\documentclass[aps,prl,twocolumn,superscriptaddress]{revtex4-1}
\usepackage{graphicx}
\usepackage{hyperref}

\hypersetup{
	colorlinks=true,
	linkcolor=blue,
	citecolor=blue,
	urlcolor=blue
}

\newcounter{firstbib}

\begin{document}
	
\title{Cavity cooling of a levitated nanosphere by coherent scattering}	
	
	\author{Uro\v s Deli\' c}
	\email{uros.delic@univie.ac.at}
	\author{Manuel Reisenbauer}
	\author{David Grass}
	\altaffiliation{Present address: Department of Chemistry, Duke University, Durham, North Carolina 27708, United States}
	\author{Nikolai Kiesel}
	\affiliation{Vienna Center for Quantum Science and Technology (VCQ), Faculty of Physics, University of Vienna, Boltzmanngasse 5, A-1090 Vienna, Austria}
	\author{Vladan Vuleti\' c}
	\affiliation{Department of Physics and Research Laboratory of Electronics, Massachusetts Institute of Technology, Cambridge, Massachusetts 02139, USA}
	\author{Markus Aspelmeyer}
	\affiliation{Vienna Center for Quantum Science and Technology (VCQ), Faculty of Physics, University of Vienna, Boltzmanngasse 5, A-1090 Vienna, Austria}

\date{\today}

\begin{abstract}
	
	%We characterize the coupling of the particle motion to the cavity mode for different tweezer polarizations and different positions along the cavity axis.
	
	We report three-dimensional cooling of a levitated nanoparticle inside an optical cavity. The cooling mechanism is provided by cavity-enhanced coherent scattering off an optical tweezer.  The observed 3D dynamics and cooling rates are as theoretically expected from the presence of both linear and quadratic terms in the interaction between the particle motion and the cavity field. By achieving nanometer-level control over the particle location we optimize the position-dependent coupling and demonstrate axial cooling by two orders of magnitude at background pressures of $6\times10^{-2}$~mbar. We also estimate a significant ($> 40$ dB) suppression of laser phase noise heating, which is a specific feature of	the coherent scattering scheme. The observed performance implies that quantum ground state cavity cooling of levitated nanoparticles can be achieved for background pressures below $1\times 10^{-7}$ mbar. 
\end{abstract}

%\keywords{}

\maketitle

Laser cooling and trapping is at the heart of modern atomic physics. In its most basic form, motional cooling of atoms \cite{Phillips1998, Metcalf1999, Cohen-Tannoudji2011,Hood2000,Domokos2001,Boozer2006} or molecules \cite{Shuman2010, Hummon2013, Lim2018, Anderegg2018, McCarron2018} is provided by the total recoil from both absorption of Doppler-shifted laser photons and the subsequent spontaneous emission. In contrast, coupling the motion of a particle to an optical cavity field can be used for cooling schemes that do not rely on the internal structure of the particle \cite{Vuletic2000,Horak1997}. This is of particular importance for increasingly complex or massive particles, for which transitions between internal energy levels become inaccessible. One highly successful method is to exploit dispersive coupling inside a driven cavity, where the position-dependent cavity frequency shift induced by the particle provides an optomechanical interaction. Demonstrations of this effect include cavity cooling of atomic systems \cite{Chan2003,Maunz2004,Nussmann2005,Fortier2007}, as well as recent experiments in cavity optomechanics that explore the quantum regime of solid state mechanical resonators \cite{Gigan2006, Arcizet2006a, Schliesser2006, Thompson2008, Teufel2011, Chan2011c,RMP2014}. For levitated nanoparticles \cite{Kiesel2013,Asenbaum2013,Millen2015,Fonseca2016}, this cooling scheme is inherently limited by the laser field driving the cavity. Specifically, large drive powers induce co-trapping by the cavity field and deteriorate cooling rates \cite{Delic2018}, while laser phase noise prohibits ground state cooling at the relevant nanoparticle trap frequencies \cite{Rabl2009,Jayich2012,Safavi-Naeini2013,SI}.

\begin{figure}
	\includegraphics[width=\linewidth]{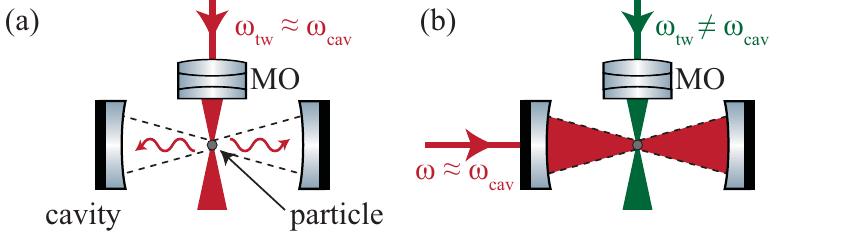}%
	\caption{\label{fig:paradigm} Different paradigms for cavity cooling of a levitated nanosphere. (a) Cavity cooling by coherent scattering from an optical tweezer is based on dipole radiation being emitted into an \textit{empty} cavity, giving the best performance for a particle placed at the intensity minimum of the cavity mode. (b) In standard dispersive optomechanics an external laser drives both the cavity and the scattering. Optimal cooling is at the largest intensity gradient of the cavity mode.}
\end{figure}

A promising alternative is cavity cooling by coherent scattering from an optical trapping field (Fig.~\ref{fig:paradigm}). In this case, a driven dipole (here: the nanosphere) produces scattering that is coherent with the drive field (here: the optical trap laser). Scattering of these photons into an initially empty cavity provides a cooling mechanism \cite{Vuletic2001}. As is usual in cavity cooling, the proper red-detuning of the drive field from the cavity allows to resonantly enhance the scattering processes that remove energy from the particle motion. Dispersive coupling schemes also originate in coherent scattering, where the drive field is the externally pumped cavity field. There, the interaction with the cavity field is determined by the scattering cross section with an independently populated cavity mode, which is typically very small for levitated nanoparticles. In contrast, in coherent scattering a photon can only enter the cavity via the scattering process that cools the particle motion. Efficient cooling does not require an additional strong intracavity field, which has the immediate advantage of lifting the limitations on drive laser power by co-trapping. 

In this Letter we demonstrate cavity cooling by coherent scattering for a levitated dielectric nanoparticle along with its unique features. We report genuine 3D cavity cooling, an effect that has thus far only been demonstrated in 1D with atoms \cite{Hosseini2017, Leibrandt2009}. By positioning the particle with $8$~nm precision along the cavity axis \cite{Delic2018} we can optimize coherent scattering rates. For a particle placed at a node of the cavity field we observe axial cooling factors beyond $100$, well described by a simple theory based on linear and quadratic optomechanical interactions. We estimate that laser phase noise of the coherently scattered radiation is suppressed by four orders of magnitude, removing a major obstacle for motional ground state cooling. 

\begin{figure}
	\includegraphics[width=\linewidth]{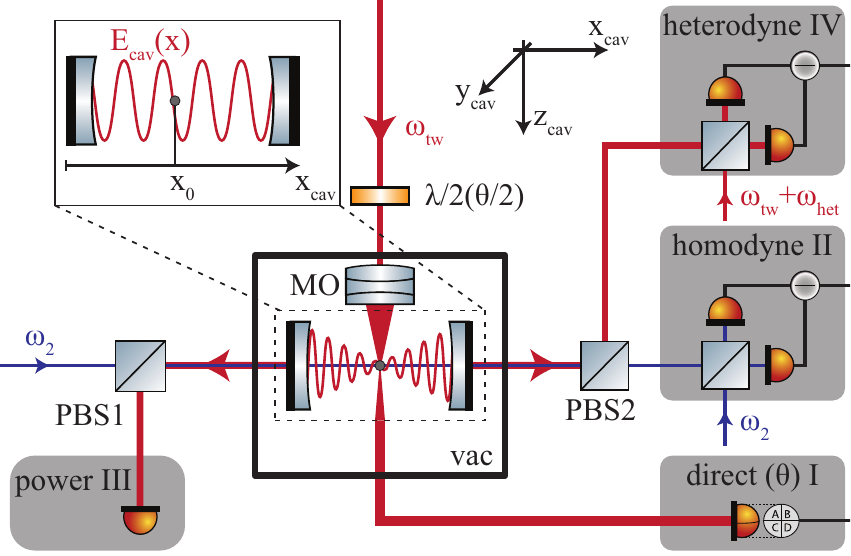}%
	\caption[Experimental setup for coherent scattering]{\label{fig:cohscattsetup}Setup for cooling by coherent scattering. An optical tweezer is formed by a laser at frequency $\omega_{tw}$ that is tightly focussed by a microscope objective (MO) inside a vacuum chamber (vac). It levitates a nanoparticle at the center of a high-finesse Fabry-P\' erot cavity. Its linear polarization is set by a half-wave plate ($\lambda/2$). A weak locking beam is derived from the tweezer laser and drives the cavity resonantly at frequency $\omega_2$, allowing $\omega_{tw}$ and $\omega_2$ to be stably locked relative to the cavity frequency. Four independent detection schemes (I-IV) monitor the particle motion and the cavity field (see main text for details; PBS: polarizing beam splitter; $\omega_{het}$: heterodyne demodulation frequency). Inset: The particle is trapped at a position $x_0$ relative to a cavity antinode. Maximal cavity cooling of the $x$-motion by coherent scattering occurs for $x_0=\lambda/4$, i.e. at a cavity node.}
\end{figure}

\textit{Theory.} Consider a nanoparticle that is trapped with an optical tweezer of waists $W_{x,y}$ inside an empty optical cavity of mode volume $V_{cav}$ (waist $w_0$) and at position $x_0$ along the cavity axis (Fig. \ref{fig:cohscattsetup}). The interaction of the induced dipole with the local electric field is then to first approximation described by the Hamiltonian:
\begin{eqnarray}
H_{dip}&=&-\frac{\alpha}{2}\left|\vec{E}_{tw}\right|^2-\frac{\alpha}{2}\left|\vec{E}_{cav}\right|^2-\alpha\Re\left(\vec{E}_{tw}\cdot\vec{E}_{cav}^*\right)\label{eq:dipoleint}\\
\vec{E}_{tw}&\approx& \frac{1}{2}\frac{\epsilon_{tw}}{\sqrt{1+(z/z_R)^2}} e^{-\frac{x^2}{W_x^2}}e^{-\frac{y^2}{W_y^2}} e^{ikz}e^{i\omega_{tw}t}\vec{e}_{y}+\text{H.c.}\nonumber\\
\vec{E}_{cav}&\approx&\epsilon_{cav}(\hat{a}^\dagger e^{-i\omega_{cav}t}+\hat{a}e^{i\omega_{cav}t})\cos k(x_0+x)\vec{e}_{y_{cav}}\nonumber
\end{eqnarray}
Here $E_{tw}$ and $E_{cav}$ are the electric fields of the tweezer and the cavity mode, respectively (with: $\epsilon_{cav}=\sqrt{\hbar\omega_{cav}/(2\varepsilon_0 V_{cav})}$ and $\epsilon_{tw}=\sqrt{4P_{tw}/(W_xW_y\pi\varepsilon_0 c)}$; tweezer frequency: $\omega_{tw}$; cavity resonance frequency: $\omega_{cav}$; particle polarizability: $\alpha$; cavity field operators: $\hat{a}^{\dagger}$ and $\hat{a}$; vacuum permittivity: $\varepsilon_0$; speed of light: $c$; wavenumber: $k$; Rayleigh length: $z_R$).

The first term corresponds to the potential energy of the particle in the optical tweezer. The second term describes the dispersive optomechanical interaction of conventional cavity optomechanics that couples the particle to the intensity distribution of the cavity field. It is maximized at cavity positions of maximal intensity gradient \cite{Romero-Isart2011,Chang2010,Nimmrichter2010,RMP2014}. The third term is the interference term between the tweezer and cavity field and represents the coherent scattering interaction \cite{Leibrandt2009,Hosseini2017}: When the tweezer frequency approaches a cavity resonance, the cavity mode density alters the emission spectrum of the dipole radiation and cavity-enhanced coherent scattering can occur \cite{Vuletic2001}. It has several unique features: First, due to the directionality of the scattered dipole radiation, the interaction strength strongly depends on the polarization of the trap laser. Coherent scattering is driving the cavity through $E_d(\theta)=\alpha\epsilon_{cav}\epsilon_{tw}\sin\theta/(2\hbar)$, where $\theta$ is the angle between the polarization vector and the cavity axis. A linearly polarized trap laser with $\theta = \pi/2$ maximizes the overlap of the dipole radiation pattern with the cavity mode. Second, the interaction scales with the local field strengths of both optical trap and cavity. For cavities with large mode volume the focused trap laser significantly boosts the interaction strength, specifically with $\epsilon_{tw}/\epsilon_{cav}\propto w_0/W_{x,y}$ when compared to dispersive coupling. Finally, the interaction is linear in the cavity electric field, which to first order yields the optomechanical interaction \cite{SI}:
\begin{eqnarray}
\frac{H_{CS}}{\hbar}&=&E_d(\theta)\cos kx_0 (\hat{a}^\dagger+\hat{a})-iE_d(\theta)k\cos kx_0(\hat{a}^\dagger-\hat{a})\hat{z}\nonumber\\
&+&E_d(\theta)k\sin kx_0 (\hat{a}^\dagger+\hat{a})(\hat{x}\sin\theta+\hat{y}\cos\theta).\label{eq:couplingHam}
\end{eqnarray} 
Here, $\hat{x}$ and $\hat{y}$ refer to the particle motion relative to the trap laser polarization. The coupling rates $g_{\{j=x,y,z\}}\propto E_d(\theta)kj_{zpf}$ formed from Eq.~\ref{eq:couplingHam} depend on polarization ($\theta$) and particle position ($x_0$). The optimal position for cavity cooling of the $x/y$-motion is at the cavity node ($|\sin kx_0|=1$), which is well known for light-atom interaction inside a standing wave \cite{Fortier2007,Russo2009,Cirac1992,Leibrandt2009} and in stark contrast to cooling via dispersive coupling of standard cavity optomechanics. Intuitively, the particle acts as an intracavity emitter. At the cavity node, i.e. the intensity minimum of the cavity standing wave, no emission can occur due to destructive interference of the scattered light. The intracavity photon number $n_{\text{phot}}=E_d^2(\theta)\cos^2kx_0/((\kappa/2)^2+(\omega_{tw}-\omega_{cav})^2)$ is accordingly zero ($\kappa$: cavity linewidth). The particle motion along the cavity axis, however, results in directional photon scattering into Doppler-shifted (Stokes and Anti-Stokes) motional sidebands, which do not interfere. As a consequence, the light scattered into the cavity will consist only of Stokes (heating) and Anti-Stokes (cooling) photons, with an imbalance in the scattering rates created by the cavity and leading to cooling of the $x$-motion \cite{Leibrandt2009}. 

On the other hand, due to the $z$-motion along the tweezer axis the particle experiences a phase-modulated drive field. In other words, the motional sidebands for the $z$-direction are already imprinted in the spectrum of our coherent emitter, with maximum emission and hence scattering rate at the cavity antinode ($|\cos kx_0|=1$). A proper choice of both particle position and tweezer polarization therefore allows to achieve genuine 3D cavity cooling.

\textit{Experiment.} The experimental setup is shown in Fig.~\ref{fig:cohscattsetup}. A microscope objective (NA $0.8$) and a near-confocal high-finesse Fabry-P\' erot cavity (Finesse $\mathcal{F} =73.000$; linewidth $\kappa = 2\pi \times 193$~kHz, length $L=1.07$~cm, waist $w_0=41.1~\mu$m, resonance frequency $\omega_{cav}$) are mounted inside a vacuum chamber. The microscope objective focuses a $1064$~nm laser (frequency $\omega_{tw}=\omega_{cav} - \Delta$, power $P_{tw}\approx 0.17$~W) to a waist of $W_x\approx 0.67~\mu$m and $W_y\approx 0.77~\mu$m, forming an optical tweezer that traps silica nanospheres (specified radius $71.5$~nm). The trap is elliptical in the transverse plane with non-degenerate mechanical frequencies $(\Omega_x,\Omega_y,\Omega_z)/2\pi=(190,170,38)$ kHz. The microscope objective is mounted on a three-axis nanopositioner with a step size of approximately $8$~nm. To control the detuning $\Delta$ between the optical trap laser and the cavity resonance frequency, a part of the trap light is frequency shifted to $\omega_2=\omega_{cav}-\text{FSR}-\Delta$ and weakly pumps the optical cavity (free spectral range $\text{FSR}=2\pi\times14$~GHz). It provides a locking signal that enables the source laser for the optical tweezer to follow the freely drifting Fabry-P\' erot cavity. The locking laser and the optical tweezer address different cavity resonances such that the mode populated via coherent scattering is initially empty. 

The experiment has four detection channels (Fig.~\ref{fig:cohscattsetup}(b)). Direct detection of the particle motion in all three directions (I) is obtained in forward scattering of the optical tweezer \cite{Gieseler2012}. Homodyne detection of the locking laser in cavity transmission (II) allows for a standard optomechanical position detection along the cavity axis. This is used to align the particle with respect to the cavity field without relying on the coherently scattered light. We also directly measure the power of the coherently scattered photons into the optical cavity (III) by monitoring the field leaking out of the left cavity mirror. Finally, a spectrally resolved characterization of these photons is enabled by a heterodyne detection of the emission from the right cavity mirror (IV). 

\begin{figure}
	\includegraphics[width=\linewidth]{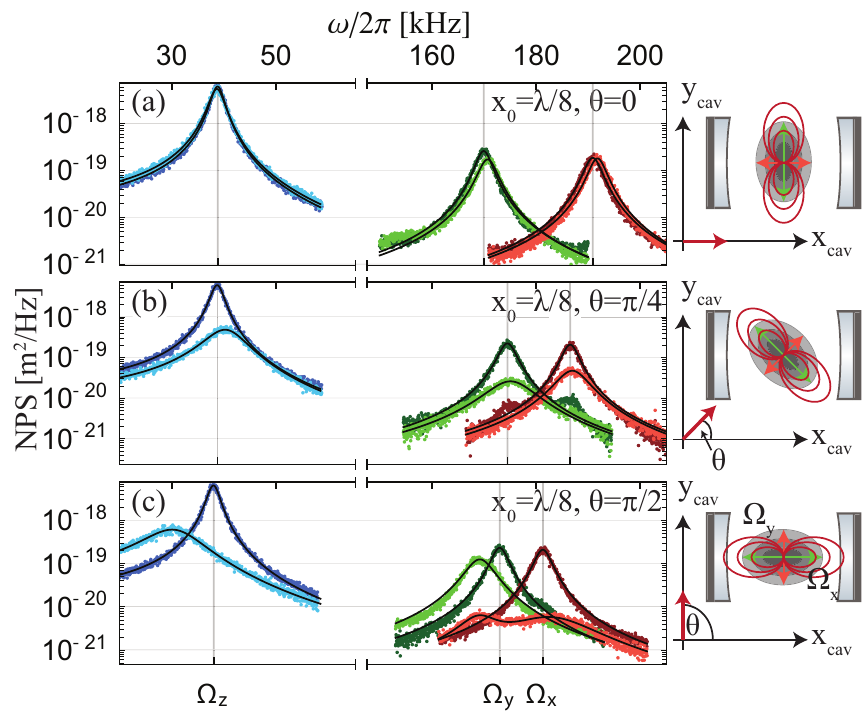}
	\caption{\label{fig:polscan} Polarization dependent 3D cavity cooling. Shown are noise power spectra (NPS) measured with direct detection (I) for a particle located at $x_0=\lambda/8$ to couple all three directions of motion and for three different tweezer polarizations as illustrated on the right panel. The red arrow indicates the polarization. The sketch also indicates the transverse optical tweezer potential (grey ellipse) and the dipole emission (red ellipses). NPS in each panel have been obtained along the tweezer axis ($z$, blue) and in its transverse directions ($x$, red; $y$, green). Cooling measurements are performed with a tweezer detuning close to the mechanical frequency ($\Delta=2\pi\times 300$~kHz, bright color). Measurements at large detuning ($\Delta=2\pi\times 4$~MHz, dark color) serve as reference for no cooling (see main text). (a) At $\theta=0$ no cooling is observed, because polarization along the cavity axis suppresses scattering into the cavity. (b) At $\theta=\pi/4$ full 3D cavity cooling by coherent scattering is observed, since the cavity axis does not coincide with a principal axis of the optical tweezer. Cooling both broadens the spectra and reduces the overall area, while the mechanical frequency is shifted due to an optical spring. (c) For $\theta=\pi/2$ scattering into the cavity is maximal, as is the cooling along the cavity axis ($x$) and the tweezer axis ($z$).}
\end{figure}

\begin{figure*}
	\includegraphics[width=\linewidth]{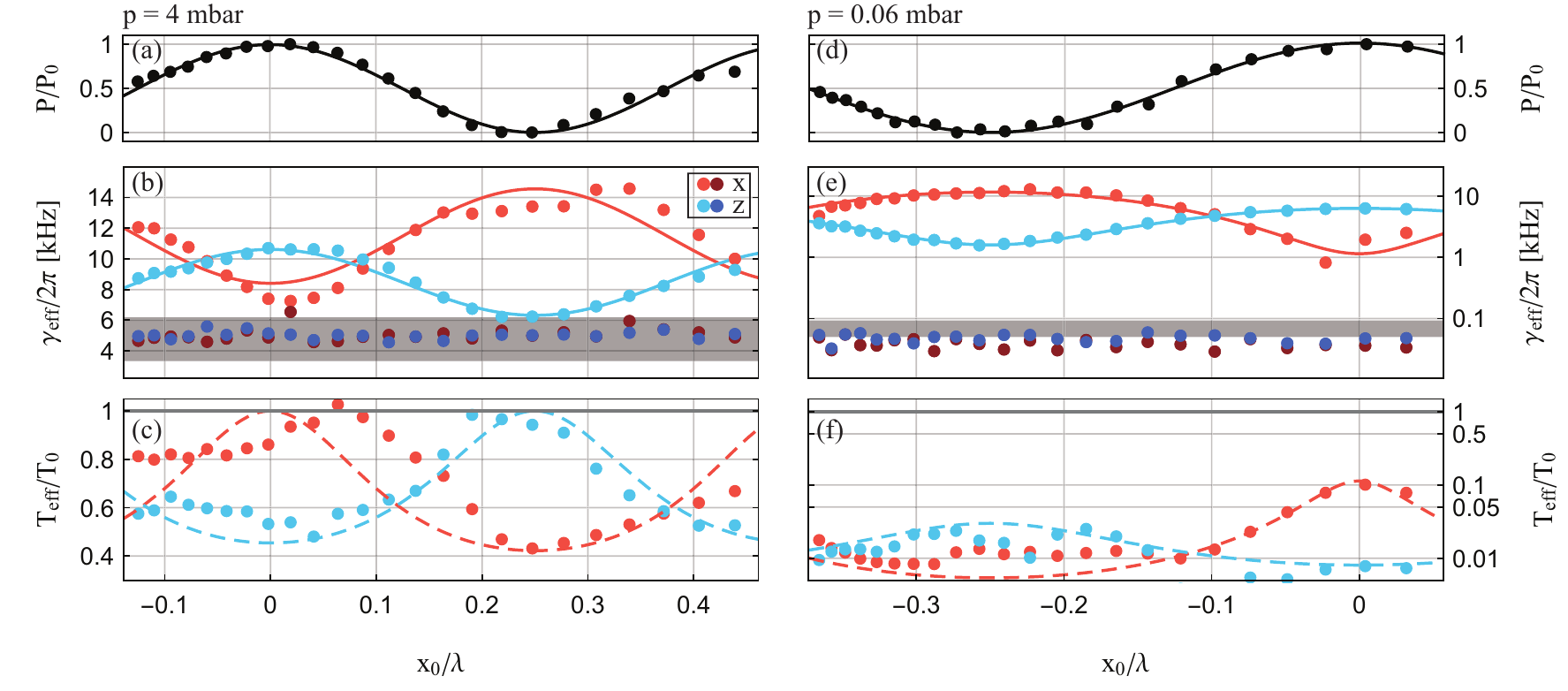}
	\caption{\label{fig:cavityscan} Position dependent cavity cooling. Shown are relative coherent scattering powers $P/P_0$ (top), mechanical damping rates $\gamma_{\text{eff}}$ (middle) and inverse cooling factors $T_{\text{eff}}/T_0$ (bottom) for different particle positions $x_0$ along the cavity axis and at background pressures of $p=4$~mbar (left) and $0.06$~mbar (right). Top panel ((a),(d)): Coherent scattering into the cavity mode. The black line is a fit to the data following the cavity standing wave. The scattering is minimal (maximal) for a particle placed at the node $x_0=\lambda/4$ (antinode $x_0=0$) of the cavity field. Middle panel ((b),(e)): The damping $\gamma_{\text{eff}}$ of the nanoparticle motion is obtained via the width (FWHM) of the NPS for the $x$-axis (red) and $z$-axis (blue). Bright colors indicate measurements with cavity cooling ($\Delta/2\pi=400$~kHz), dark colors without cooling ($\Delta/2\pi=4$~MHz). The grey line shows the theoretically predicted gas damping $\gamma_{gas}$, which agrees with the damping observed in the absence of cooling. As expected, maximal damping along the $x$ ($z$)-direction is obtained for minimal (maximal) coherent scattering powers at $x_0=\lambda/4$ ($x_0=0$), as predicted by our theoretical model (solid line; see main text). Bottom panel ((c),(f)): The effective mode temperatures $T_{\text{eff}}$ are obtained by NPS integration (see main text). As expected for both directions, maximum damping implies maximal cooling. Purely linear coupling would result in a maximum temperature of $T_{\text{eff}}/T_0=1$ (grey line). A theoretical model that also includes quadratic coupling matches the data very well without free parameters (dashed lines).
	}
\end{figure*}

\textit{Polarization dependent cavity cooling.} The effect of cavity-enhanced coherent scattering depends on the polarization of the optical tweezer. We investigate cooling by coherent scattering for three linear polarization angles $\theta=0$, $\theta=\pi/4$ and $\theta=\pi/2$. We record the particle motion using direct detection (Fig.~\ref{fig:cohscattsetup}, I). For these measurements the particle is positioned at the maximum intensity gradient of the empty cavity mode ($x_0=\lambda/8$) \cite{SI} such that cooling by coherent scattering affects all motional axes. For each polarization we compare the cooled motion obtained at a trap laser detuning $\Delta/2\pi=300~$kHz to an uncooled motion obtained at a far-detuning $\Delta/2\pi=4$~MHz. 

Initially, we set the trap laser polarization along the cavity axis ($\theta=0$) by minimizing the scattering into the empty cavity mode (Fig. \ref{fig:polscan}(a)). For perfect polarization alignment a complete suppression of this scattering would be expected. We achieve a suppression by a factor of $100$, limited by the alignment between tweezer and cavity axes \cite{SI}. The resulting coherent scattering is responsible for modest cavity cooling of the $y$- and $z$-motion. For $\theta=\pi/4$ (Fig. \ref{fig:polscan}(b)) all directions of motion are coupled to the cavity mode with rates $g_j/2\pi=(20,30,71)~$kHz and we observe genuine 3D cooling by coherent scattering. Rotating the polarization to $\theta=\pi/2$ (Fig. \ref{fig:polscan}(c)) optimizes cooling of the $x$- and $z$-motion. Cooling of the $y$-motion is explained by a slightly elliptical trap polarization, with inferred coupling rates $g_j/2\pi=(42,16,94)~$kHz. 

\textit{Position dependent cavity cooling.} We set the polarization angle $\theta=\pi/2$ to maximize the scattering into the cavity mode. The cooling performance is now measured at a detuning of $\Delta/2\pi=400$~kHz. We move the particle in steps of $\sim 20$~nm along the cavity axis at pressures of $p=4$ mbar (Fig.~\ref{fig:cavityscan}(a)-(c)) and $p=0.06$~mbar (Fig.~\ref{fig:cavityscan}(d)-(f)). The particle position is deduced from the scattered power (detector III, Fig.~\ref{fig:cavityscan}(a) and (d)) and independently confirmed by the homodyne (II) and heterodyne detection (IV) \cite{SI}. The maximal effective damping $\gamma_{\text{eff}}^{x}$ ($\gamma_{\text{eff}}^{z}$) of the particle motion is observed at the cavity node (antinode), in agreement with theory (Fig.~\ref{fig:cavityscan}(b) and (e)). This is a unique signature of cooling by coherent scattering. We fit the mechanical damping by a simple model $\gamma_{\text{eff}}^{x[z]}=\gamma_{min}+(\gamma_{max}-\gamma_{min})\sin^2 kx_0[\cos^2 kx_0]$, yielding the optical linear damping rate $(\gamma_{max}^{x[z]}-\gamma_{gas})/2\pi=10 [6.2]$~kHz. From this we are able to extract the maximal coupling rates $g_x=2\pi\times 60$~kHz and $g_z=2\pi\times 120$~kHz for the respective optimal particle positions, yielding a cavity drive $E_d/2\pi= 2.5\times 10^9$~Hz. For comparison, the cavity drive required to reach the same coupling rate $g_x$ in the dispersive regime is $E_d^{disp}/2\pi=4.2\times 10^{10}~$Hz, which corresponds to an intracavity photon number that is larger by a factor of $(E_d^{disp}/E_d)^2\approx 280$.  The position dependent coupling of coherent scattering provides an additional suppression of $n_{\text{phot}}$. At the optimal position for axial coupling, i.e. in the proximity of the cavity node, we observe a reduction of $n_{\text{phot}}$ by a factor of $\sim50$ (Figure \ref{fig:cavityscan}(a), (d)). As a direct consequence, our coherent scattering scheme suppresses phase noise heating of the particle motion by a factor of $1.4\times 10^4$ compared to a driven cavity. In a 3D cooling configuration the suppression factor is still on the order of $60$ \cite{SI}.

We obtain the effective mode temperatures of the $x$- and $z$-motion $T_{\text{eff}}^x$ and $T_{\text{eff}}^z$ from the area underneath the noise power spectra and normalized to the bath temperature $T_0$ measured without cooling (Fig.~\ref{fig:cavityscan}(c) and (f)). At $p=0.06$ mbar we observe temperatures below $T_0$ even where no cooling is expected according with the model discussed so far. For the $x$-motion, including a quadratic interaction with an average temperature $T_{\text{eff}}^x/T_0^x|_{\text{quad}}=0.11$ \cite{SI,Nunnenkamp2010} yields good agreement with the experimental data. The strong cooling of the $z$-motion is mostly due to a small angle between the tweezer axis and the $z_{cav}$-axis, resulting in a projection of the $z$-motion onto the cavity axis. For comparison, the dashed line in Fig.~\ref{fig:cavityscan}(c) and (f) is based on a theoretical model that includes the linear and, in case of the $x$-motion, quadratic interaction \cite{SI}.

\textit{Conclusion.} We have conducted a systematic experimental study of cavity cooling by coherent scattering and demonstrated genuine 3D cavity cooling, making cavity cooling self-sufficient for experiments in ultra-high vacuum. Maximizing the cooling along the cavity axis, we obtain coupling rates of $g_x = 2\pi\times60$ kHz and $g_z=2\pi\times 120$~kHz. The position of optimal axial cooling comes with more than $4$ orders of magnitude suppression of laser phase noise heating, thus removing the major obstacles for motional ground state cooling in levitated cavity optomechanics. Currently, we achieve a minimal temperature of $T_{\text{eff}}^{x[z]}\approx 1$~K, mainly limited by the modest vacuum pressure of $p =6\times10^{-2}$~mbar. Given our sideband resolution we expect an axial phonon number of $\bar{n}_x^{min}=(\kappa/(4\Omega_x))^2+\kappa\Gamma_{\text{rec}}/(4g_x^2)=0.16$ when operating the experiment in the recoil-limited regime ($p\approx 10^{-7}$~mbar) \cite{SI,Jain2016}. As a new method for levitated particles, the coherent scattering as presented here can enable still stronger coupling rates using higher power in the optical tweezer and larger particles. This opens the path to the regime of ultra-strong coupling where the coupling rate exceeds both mechanical frequency and cavity decay rate, giving rise to novel quantum optomechanical effects \cite{Genes2008Entanglement,HoferPhD}.

\begin{acknowledgments}
\textit{Acknowledgments.} This project was supported by the European Research Council (ERC CoG QLev4G), by the ERA-NET programme QuantERA, QuaSeRT (Project No. 11299191; via the EC, the Austrian ministries BMDW and BMBWF and research promotion agency FFG), by the Austrian Science Fund (FWF, Project TheLO, AY0095221, START) and the doctoral school CoQuS (Project W1210), and the Austrian Marshall Plan Foundation.

We thank Helmut Ritsch, Pablo Solano and Lorenzo Magrini for valuable discussions. We thank the team of Martin Weitz (Tobias Damm) for their help with cutting the cavity mirrors.
\end{acknowledgments}

\textit{Note.} We recently became aware of similar work done by Windey et al. \cite{Windey2018} and Gonzalez-Ballestero et al. \cite{Gonzalez2019}.


%\bibliographystyle{apsrev4-1}
%\bibliography{references}

\setcounter{figure}{0}
\setcounter{equation}{0}
\renewcommand{\thefigure}{S\arabic{figure}}
\renewcommand{\theequation}{S\arabic{equation}} 

\newpage

\begin{widetext}
	
\section{Supplemental material}

\subsection{Theory}

Assuming that the total electric field is a sum of the tweezer field and the subsequently generated cavity field, the Hamiltonian of the dipole interaction from the main text is:
\begin{equation}
\hat{H}_{dip}=-\frac{1}{2}\alpha\left|\vec{E}_{cav}+\vec{E}_{tw}\right|^2=-\frac{1}{2}\alpha|\vec{E}_{tw}|^2-\frac{1}{2}\alpha|\vec{E}_{cav}|^2-\alpha\Re \left(\vec{E}_{tw}\vec{E}_{cav}^*\right),\label{Hamiltonian}
\end{equation}
where $\alpha=3\varepsilon_0 V\frac{\varepsilon_n-1}{\varepsilon_n+2}$ is the polarizability of a nanosphere with volume $V$ and relative dielectric permittivity $\varepsilon_n$. The electric fields are given as:
\begin{eqnarray}
\vec{E}_{tw}&=&\frac{1}{2}\epsilon_{tw}\frac{1}{\sqrt{1+\left(\frac{z}{z_R}\right)^2}}e^{-\frac{x^2}{W_x^2}}e^{-\frac{y^2}{W_y^2}}e^{ikz}e^{i\varphi_G(z)}e^{i\omega_{tw}t}\vec{e}_{y}+\text{H.c.},\hspace{0.5cm}\epsilon_{tw}=\sqrt{\frac{4P_{tw}}{W_xW_y\pi\varepsilon_0 c}}\nonumber\\
\vec{E}_{cav}&=&\epsilon_{cav}\cos k(x_0+x)(\hat{a}^\dagger e^{-i\omega_{cav}t}+\hat{a}e^{i\omega_{cav}t})\vec{e}_{y_{cav}},\hspace{0.5cm}\epsilon_{cav}=\sqrt{\frac{\hbar\omega_{cav}}{2\varepsilon_0V_{cav}}}
\end{eqnarray}
where $\varphi_G(z)=-\arctan(z/z_R)$ is the Gouy phase of the tweezer electric field with the Rayleigh range $z_R=W_xW_y\pi/\lambda$ and the waists of the elliptical tweezer focus $W_x$ and $W_y$, $P_{tw}$ is the tweezer power, $\omega_{tw}$ and $\omega_{cav}$ are the tweezer frequency and the cavity resonant frequency, $V_{cav}=w_0^2\pi L/4$ is the cavity mode volume with cavity waist $w_0=41.1~\mu\text{m}$ and cavity length $L=1.07~$cm. The nanosphere is positioned on the cavity axis at an arbitrary position $x_0$ with respect to an antinode.

The first term in Equation \ref{Hamiltonian} is the three-dimensional harmonic potential for the nanosphere. From the mechanical frequencies $(\Omega_x,\Omega_y,\Omega_z)/2\pi=(190,170,38)$~kHz we estimate the focal power $P_{tw}\approx 0.17$~W and the waists $W_x=0.67~\mu\text{m}$ and $W_y=0.77~\mu\text{m}$ \cite{NovotnyBook}. The second term is the standard interaction with the intensity of the cavity mode, which is now driven with the coherently scattered light. Note that no additional drive through the cavity mirrors is assumed. The third term $H_{CS}=-\alpha\Re(\vec{E}_{tw}\vec{E}^*_{cav})$ is the interference between the tweezer and the cavity electric field, which can be switched off for the orthogonally polarized tweezer and cavity modes. In our experiment, it's possible to have almost a perfect overlap of the two fields, as both the tweezer and the driven cavity mode can be polarized along the $y_{cav}$-axis. In general $\vec{E}_{tw}\vec{E}^*_{cav}=E_{tw}E_{cav}^*\sin\theta$, where $\theta$ is the angle between the polarization vector of the tweezer electric field and the cavity axis $x_{cav}$ (See Fig. 3 in the main text). The $x$-$y$ oscillation plane of the tweezer potential follows the rotation of the tweezer polarization, such that the transverse motion is projected onto the cavity axis as $x\rightarrow x\sin\theta+y\cos\theta$.

Due to a rotating wave approximation, the fast oscillating terms in the fields interference can be omitted. In another words, the scattering process annihilates a photon in the tweezer mode and creates a photon in the cavity mode, while the process of annihilating two photons is suppressed. Therefore, the most general expression of the interaction Hamiltonian is:
\begin{eqnarray}
\frac{H_{CS}}{\hbar}&=& E_d(\theta)\left(\hat{a}^\dagger e^{i(kz-\varphi_G(z))}+\hat{a}e^{-i(kz-\varphi_G(z))}\right)\cos k(x_0+x\sin\theta+y\cos\theta)\label{HamCS1}\\
&=&-(\hat{a}^\dagger+\hat{a})\left[\underbrace{E_d(\theta)kx_{zpf}\sin\theta\sin kx_0}_{g_x(\theta,x_0)}\frac{\hat{x}}{x_{zpf}}+\underbrace{E_d(\theta)ky_{zpf}\cos\theta\sin kx_0}_{g_y(\theta,x_0)}\frac{\hat{y}}{y_{zpf}}\right]+\nonumber\\
&+&i(\hat{a}^\dagger-\hat{a})\underbrace{E_d(\theta)(k-1/z_R)z_{zpf}\cos kx_0}_{g_z(\theta,x_0)} \frac{\hat{z}}{z_{zpf}}+E_d(\theta)\cos kx_0(\hat{a}^\dagger+\hat{a}),\label{HamCS}
\end{eqnarray}
where we define the cavity drive as $E_d(\theta)=\alpha\epsilon_{tw}\epsilon_{cav}\sin\theta/(2\hbar)$. For a silica nanosphere with a nominal radius of $r=71.5~$nm and permittivity $\varepsilon_n\approx 2.1$, we calculate the expected drive to be $E_d(\pi/2)/2\pi\approx 2.8\times 10^9$~Hz, which is close to the determined $E_d/2\pi=2.5\times 10^9~$Hz from the measurements in the main text. In deriving Eq. \ref{HamCS} from Eq. \ref{HamCS1} we looked into the following contributions:
\begin{itemize}
	\item \textbf{Cavity drive:} $E_d(\theta)\cos kx_0(\hat{a}^\dagger+\hat{a})$ describes how the coherently scattered light off the nanosphere drives the cavity mode. The maximum scattering into the cavity mode is for $\theta=\pi/2$, when the coherently scattered light shares the polarization of the driven cavity mode. The cavity enhances the scattered light with a maximum intracavity photon number reached for a nanosphere positioned at the cavity antinode:
	\begin{equation}
	n_{phot}=\frac{E_d^2(\theta)\cos^2 kx_0}{\left(\frac{\kappa}{2}\right)^2+\Delta^2},
	\end{equation}
	where $\kappa$ is the cavity linewidth and $\Delta$ is the tweezer detuning with respect to the cavity resonance. The same result is obtained from the mode overlap of the dipole radiation pattern and the cavity electric field \cite{Tanji-Suzuki2011,Motsch2010}.
	
	\item \textbf{Coupling to the $z$-motion:} The nanosphere is in the Lamb-Dicke regime as the nanosphere motion is significantly smaller than the laser wavelength ($k\sqrt{\langle z^2\rangle}\ll 1$). Therefore, the phase of the tweezer electric field to second order is approximately $\exp(i(kz-\arctan(z/z_R)))\approx 1+i(k-1/z_R)z-(k-1/z_R)^2z^2/2$. The contribution from the Gouy phase is a factor of $kz_R\approx 9$ times smaller than the main contribution. The coupling to the $z$-motion is $g_z(\theta,x_0)=E_d(\theta)(k-1/z_R)z_{zpf}\cos kx_0$ and is maximal for a nanosphere positioned at the cavity antinode ($|\cos kx_0|=1$). We calculate the expected coupling rate $g_z(\pi/2,0)/2\pi\approx 131$ kHz.
	
	\item \textbf{Coupling to the $x$- and $y$-motion:} Linear coupling to the $x$- and $y$-motion is featured in the Taylor expansion of the cavity electric field profile:
	\begin{equation}
	\cos k(x_0+x\sin\theta+y\cos\theta)\approx \cos kx_0\left(1-\frac{k^2(x\sin\theta+y\cos\theta)^2}{2}\right)-\sin kx_0\times k(x\sin\theta+y\cos\theta).
	\end{equation}
	The linear interaction to the $x$- and $y$-motion is maximum for a nanosphere positioned at the cavity node ($|\sin kx_0|=1$), while the quadratic interaction is maximum at the cavity antinode ($|\cos kx_0|=1$). The calculated maximum linear coupling rate to the $x$-motion is $g_x(\pi/2,\lambda/4)/2\pi\approx 67$~kHz. The dispersive coupling rate achievable in the same setup with an equal cavity drive applied through a cavity mirror would be $g_x^{disp}=g_0E_d/\sqrt{(\kappa/2)^2+\Omega_x^2}\approx2\pi\times 4~$kHz  \cite{Delicb2018}, significantly smaller compared to the coherent scattering scheme.
\end{itemize}

\subsubsection{Residual coupling due to a tilt of the tweezer}

The angle between the tweezer axis and the cavity axes is $90^\circ-\varphi$, where $\varphi< 10^\circ$ is a small deviation \cite{Delicb2018}. There are two important effects due to the existence of this deviation:
\begin{itemize}
	\item The scattering into the cavity is never fully suppressed as the residual cavity drive is $E_d(\varphi)$. We measure the suppression in the following text.
	\item The $x-z$ oscillation plane is rotated by $\varphi$ with respect to the $x_{cav}-z_{cav}$ plane defined by the cavity, leading to a small coupling of the $z$-motion at the cavity node:
	\begin{equation}
	\hat{x}_{cav}=x\sin(\pi/2-\varphi)+z\cos(\pi/2-\varphi)\approx x-z\sin\varphi.
	\end{equation}
	The total linear coupling to the $z$-motion in this configuration is:
	\begin{equation}
	\bar{g}_z(\theta,x_0)=E_d(\theta)kz_{zpf}\cos kx_0-E_d(\theta)kz_{zpf}\sin\theta\sin\varphi\sin kx_0,
	\end{equation}
	which would explain the observed $z$-cooling at any point along the cavity axis in Fig. 4 in the main text. Using $\varphi\approx 6.3^\circ$ determined from the homodyne measurement, we estimate the added coupling rate to maximally be $E_d(\pi/2)kz_{zpf}\sin\varphi=2\pi\times 14$~kHz.
	
\end{itemize}

\subsubsection{Cavity cooling of the $x$- and $z$-motion}

We set the polarization $\theta=\pi/2$. We focus only on the linear interaction with the $x$- and $z$-motion in the Langevin equations:
\begin{eqnarray}
\dot{\hat{p}}_x&=&-m\Omega_x^2\hat{x}-\gamma_m\hat{p}_x-\hbar \frac{g_x(\pi/2,x_0)}{x_{zpf}} \left(\hat{a}^\dagger+\hat{a}\right)+F_{th}^x(t),\hspace{0.5cm}\dot{\hat{p}}_z=-m\Omega_z^2\hat{z}-\gamma_m\hat{p}_z-i\hbar \frac{g_z(\pi/2,x_0)}{z_{zpf}} \left(\hat{a}^\dagger-\hat{a}\right)+F_{th}^z(t)\nonumber\\
\dot{\hat{x}}&=&\frac{\hat{p}_x}{m},\hspace{7.75cm}\dot{\hat{z}}=\frac{\hat{p}_z}{m}\nonumber\\
\dot{\hat{a}}&=&-\left(\frac{\kappa}{2}+i\Delta'\right)\hat{a}+iE_d\cos  kx_0-i\frac{g_x(\pi/2,x_0)}{x_{zpf}} ~\hat{x}-\frac{g_z(\pi/2,x_0)}{z_{zpf}} ~\hat{z}+\sqrt{\kappa_{nano}(x_0)}\hat{a}_{tw}+\sqrt{\kappa_{in}}\left(\hat{a}^1_{\text{IN}}+\hat{a}^2_{\text{IN}}\right),\label{LangCoherent}
\end{eqnarray}
where $\kappa_{nano}(x_0)=4\left|\frac{k\alpha}{\varepsilon_0 w_0^2\pi}\right|^2\Delta\nu_{FSR}\cos^2 kx_0$ is the cavity input rate due to the light scattering, while $\kappa_{in}$ is the loss rate of the two cavity mirrors. The cavity is driven by the coherently scattered light off the nanosphere with a photon rate $E_d\cos kx_0$. As a result, the cavity operators include a coherent amplitude $\alpha_0$ as $\hat{a}\rightarrow\alpha_0+\hat{a}$, which is determined from the Langevin equations above:
\begin{equation}
\alpha_0(x_0)=\frac{iE_d\cos kx_0}{\frac{\kappa}{2}+i\Delta},\hspace{1cm} n_{\text{phot}}=|\alpha_0|^2. \label{cohLightAmp}
\end{equation}

After the operator displacement and only up to first order in the operators, the Langevin equations become:
\begin{eqnarray}
\hat{a}&=&-\left(\frac{\kappa}{2}+i\Delta'\right)\hat{a}-i\frac{g_x(\pi/2,x_0)}{x_{zpf}}\hat{x}-\frac{g_z(\pi/2,x_0)}{z_{zpf}}\hat{z}+\sqrt{\kappa_{nano}}\hat{a}_{tw}+\sqrt{\kappa_{in}}\left(\hat{a}^1_{\text{IN}}+\hat{a}^2_{\text{IN}}\right)\nonumber\\
\ddot{\hat{x}}&=&-\gamma_m\dot{\hat{x}}-\Omega_x^2\hat{x}-\frac{\hbar g_x(\pi/2,x_0)}{mx_{zpf}}\left(\hat{a}^\dagger+\hat{a}\right)+f_{th}(t)\nonumber\\
\ddot{\hat{z}}&=&-\gamma_m\dot{\hat{z}}-\Omega_z^2\hat{z}-i\frac{\hbar g_z(\pi/2,x_0)}{mz_{zpf}}\left(\hat{a}^\dagger-\hat{a}\right)+f_{th}(t).
\end{eqnarray}
The procedure to solve the Langevin equations for cooling of one-dimensional motion is explained in detail in \cite{Genes2008}. Note that due to the $x$- and $z$-motion being coupled to two orthogonal quadratures of the cavity field, the equations can be solved independently for the two motions. In conclusion, for a tweezer red-detuned with respect to the cavity resonance, the particle $x$- and $z$- motion will be cooled with rates depending on the particle position. 

%Due to large coupling rates, the increased mechanical damping will be comparable to the mechanical frequency, hence no approximation of the optical spring and optical damping effects can be assumed.

\subsubsection{Cavity cooling of the motion in the transverse plane of the tweezer}

A rotation of the tweezer polarization by an angle $\theta$ leads to a rotation of the trapping potential by the same angle $\theta$. We define the motion along the transverse potential semi-major and semi-minor axes as $x(t)$ and $y(t)$ with the unchanged mechanical frequencies $\Omega_x$ and $\Omega_y$, respectively. The projections of the motion onto the cavity $x_{cav}$- and $y_{cav}$-axis (defined by the cavity in case $\theta=0$):
\begin{equation}
x_{cav}=x\cos\theta+y\sin\theta,\hspace{0.5cm}y_{cav}=x\sin\theta-y\cos\theta.
\end{equation}
Let's assume the optimal position of $\sin kx_0=1$ for the cavity cooling of the motion along the $x_{cav}$-axis and the polarization angle $\theta=\pi/4$. The Hamiltonian of the interaction with the $u$- and $v$-motion projected onto the cavity axis is:
\begin{equation}
\hat{H}_{x-y_{cav}}=\hbar E_d\left(\frac{\pi}{4}\right) k\frac{\hat{x}+\hat{y}}{\sqrt{2}}\left(\hat{a}^\dagger+\hat{a}\right),
\end{equation}
with the system dynamics described by the following Langevin equations:
\begin{eqnarray}
\ddot{\hat{x}}+\gamma_m\dot{\hat{x}}+\Omega_x^2\hat{x}-\frac{\hbar E_d\left(\frac{\pi}{4}\right) k}{\sqrt{2}m}\left(\hat{a}^\dagger+\hat{a}\right)&=&f_{th}^x\nonumber\\
\ddot{\hat{y}}+\gamma_m\dot{\hat{y}}+\Omega_y^2\hat{y}-\frac{\hbar E_d\left(\frac{\pi}{4}\right)k}{\sqrt{2}m}\left(\hat{a}^\dagger+\hat{a}\right)&=&f_{th}^y\nonumber\\
\dot{\hat{a}}+\left(\frac{\kappa}{2}+i\Delta\right)\hat{a}-\frac{i}{\sqrt{2}}E_d\left(\frac{\pi}{4}\right)k(\hat{x}+\hat{y})&\approx&0.
\end{eqnarray}
The sum and the difference of the first two equations:
\begin{eqnarray}
\overbrace{(\ddot{\hat{x}}+\ddot{\hat{y}})}^{\ddot{x}_{cav}}+\gamma_m\overbrace{(\dot{\hat{x}}+\dot{\hat{y}})}^{\dot{x}_{cav}}+(\Omega_x^2\hat{x}+\Omega_y^2\hat{y})-2\frac{\hbar E_d(\frac{\pi}{4}) k}{\sqrt{2}m}\left(\hat{a}^\dagger+\hat{a}\right)&=&f_{th}^x+f_{th}^y\nonumber\\
\overbrace{(\ddot{\hat{x}}-\ddot{\hat{y}})}^{\ddot{y}_{cav}}+\gamma_m\overbrace{(\dot{\hat{x}}-\dot{\hat{y}})}^{\dot{y}_{cav}}+(\Omega_x^2\hat{x}-\Omega_y^2\hat{y})&=&f_{th}^x-f_{th}^y\nonumber
\end{eqnarray}
shows that the two-dimensional cooling of both motions is possible only in the case of non-degenerate frequencies $\Omega_x\neq\Omega_y$. Otherwise, the difference shows that the projected dynamics along the $y_{cav}$-axis would be uninfluenced by the cavity mode.

\subsubsection{Phase noise}

The classical phase and intensity noise can influence the lowest reachable phonon occupation in cavity cooling setups \cite{Safavi-Naeinib2013, Rablb2009,Jayichb2012}. In essence, due to a non-zero detuning of the cooling laser, phase noise is converted into the amplitude and intensity noise in the optomechanical cavity. Phase noise can be implemented into our calculus as a phase variation of the driving field $E_d\rightarrow E_d e^{i\phi(t)}\approx E_d(1+i\phi(t))$, further impacting the particle motion. Phase noise contribution to the minimum phonon occupation of the $x$-motion is:
\begin{equation}
\bar{n}_x^{phase}=\frac{n_{\text{phot}}}{\kappa}S_{\dot{\phi}\dot{\phi}}(\Omega_x)=\frac{E_d^2\cos^2 kx_0}{\kappa\left(\left(\frac{\kappa}{2}\right)^2+\Omega_x^2\right)}S_{\dot{\phi}\dot{\phi}}(\Omega_x),
\end{equation}
where $S_{\dot{\phi}\dot{\phi}}$ is the intrinsic laser frequency noise. Note that the added occupation due to the phase noise heating is essentially zero at the cavity node, i.e. at the position where the maximum cooling of the $x$-motion occurs. In reality, it depends on how precise we can position the nanosphere in the vicinity of the cavity node.

\subsubsection{Optomechanical cooperativity and minimum phonon occupation}

At sufficiently low pressures ($p<10^{-7}$~mbar), heating of the nanosphere motion is given by the recoil heating of the trapping laser \cite{Jainb2016}:
\begin{equation}
\Gamma_{rec,x}^{tw}=\frac{4}{5}\frac{\omega_c}{\Omega_x}\frac{I_{tw}}{mc^2}\frac{k^4|\alpha|^2}{6\pi\varepsilon_0^2}=\frac{2}{15}\frac{k^2w_0^2}{\Delta\nu_{FSR}}\underbrace{E_d^2k^2x_{zpf}^2}_{g_x^2},
\end{equation}
where $I_{tw}$ is the trapping laser intensity and $\Delta\nu_{FSR}=14~$GHz is the cavity free spectral range. As it turns out, the optomechanical cooperativity in the recoil heating limit $C_Q=4g_x^2/\kappa\Gamma_{rec,x}^{tw}$ depends only on the cavity finesse $\mathcal{F}$ and the waist $w_0$:
\begin{eqnarray}
C_Q&=&\frac{30\mathcal{F}/\pi}{k^2w_0^2}.
\end{eqnarray}
Already for the current cavity parameters ($\mathcal{F}=73,000$, $w_0=41.1\mu \textrm{m}$) we obtain $C_Q\approx 12$, a significant improvement over the cooperativity reached in the dispersive regime \cite{Delicb2018}. The minimum phonon occupation of the nanosphere $x$-motion is reached for a nanosphere placed at the cavity node ($|\sin kx_0|=1$):
\begin{equation}
\bar{n}_x=\underbrace{\left(\frac{\kappa}{4\Omega_x}\right)^2}_{\approx 0.07}+\underbrace{\frac{\Gamma_{rec,x}^{tw}\kappa}{4g_x^2}}_{\approx 0.09} +\underbrace{\bar{n}_x^{phase}}_{=0}\approx 0.16.
\end{equation}
The respective ground state occupation probability of the $x$-motion is $87\%$. 

%On the other hand, the minimum phonon occupation of the $z$-motion at the cavity antinode is limited mostly by the phase noise $S_{\dot{\phi}\dot{\phi}}(\Omega_z/2\pi)\approx 1 \text{Hz}^2/\text{Hz}$ (Mephisto data sheet):
%\begin{equation}
%\bar{n}_z=\underbrace{\left(\frac{\kappa}{4\Omega_z}\right)^2}_{\approx 1.46}+\underbrace{\frac{\Gamma_{rec,z}^{tw}\kappa}{4g_z^2}}_{\approx 0.09} +\underbrace{\bar{n}_z^{phase}}_{\approx 750.2}\approx 752.
%\end{equation}
%In the recoil heating limit ($\Gamma_{rec}^{tw}>\gamma_m n_{th}$), the occupation is $\bar{n}_x\approx 0.16$.
%where $n_{th}=k_B T_0/(\hbar\Omega_x)$ is the thermal phonon occupation at the room temperature $T_0$ and $\bar{n}_{phase}\approx 0$

\subsection{Suppression of scattering by polarization}
\begin{figure}[h!]
	\includegraphics[width=0.5\linewidth]{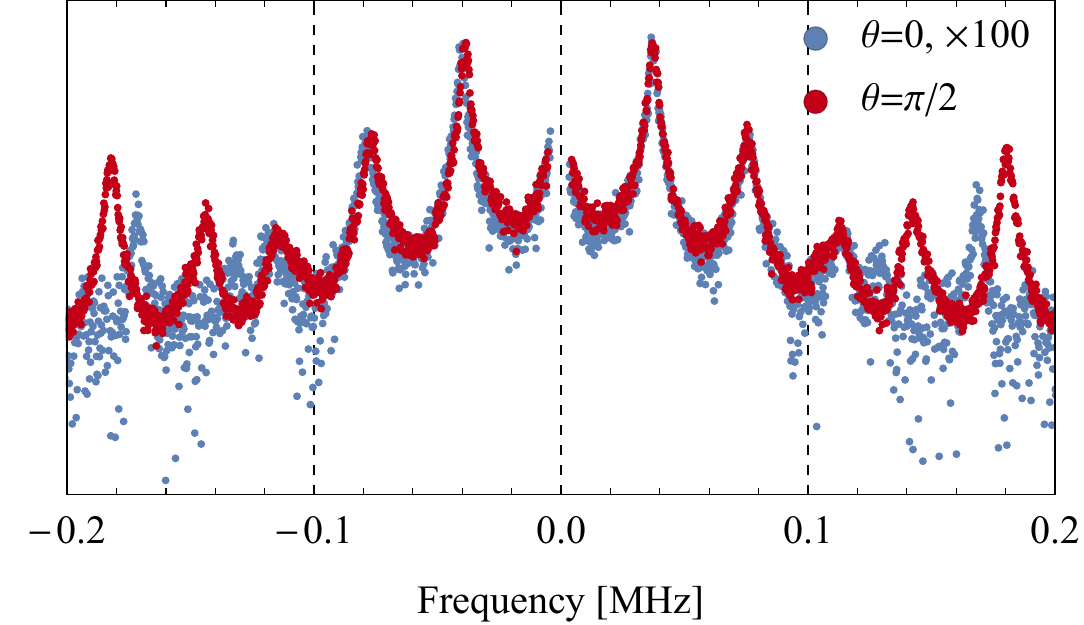}
	\caption{\label{fig:scattsupp} Overlapped heterodyne measurements for trap laser polarization $\theta=0$ and $\theta=\pi/2$. Heterodyne measurements are acquired for trap laser far detuned by $\Delta=2\pi\times 4$ MHz to avoid an affecting the particle motion. Particle is positioned halfway between a cavity node and an antinode ($x_0=\lambda/8$). The heterodyne spectrum in the case of $\theta=0$ has been multiplied by a factor of $100$ to overlap it with the case of $\theta=\pi/2$. Note that, due to the rotation of the trap axes, we couple $x$-motion and $y$-motion for $\theta=\pi/2$ and $\theta=0$, respectively.}
\end{figure}

We observe coupling of both $x$- and $z$- motion in the homodyne detection of the locking laser  (local oscillator power $0.2$ mW), which is due to a non-straight angle $90^\circ-\varphi$ between the tweezer and the cavity axis \cite{Delicb2018}. When we set the trap laser polarization $\theta=0$, the resulting angle between the polarization and the cavity axis is $\varphi$. Therefore, the residual scattering into the cavity mode is suppressed by a factor of $\sin^2\varphi$ compared to the case when $\theta=90^\circ$. We are able to directly observe the magnitude of suppression of coherent scattering by polarization with the heterodyne detection (local oscillator of $0.8$ mW power and a detuning $\omega_{het}/2\pi=21.4$ MHz from the optical tweezer frequency). We detune the tweezer by $\Delta=2\pi\times 4$~MHz to avoid affecting the particle motion. By comparing the heterodyne spectra for maximum ($\theta=90^\circ$) and minimum scattering ($\theta=0^\circ$) into the cavity mode (Fig. \ref{fig:scattsupp}), the number of scattered photons is decreased by a factor of $\sim 100$, from which we calculate the angle $\varphi\approx 5.7^\circ$. From the ratio of the overall transduction factors in the homodyne detection we obtain a similar value $\varphi\approx 6.3^\circ$, confirming that the seen suppression is consistent with the non-orthogonal tweezer and cavity axes.

\subsection{Particle positioning}

\begin{figure}[h!]
	\includegraphics[width=0.7\linewidth]{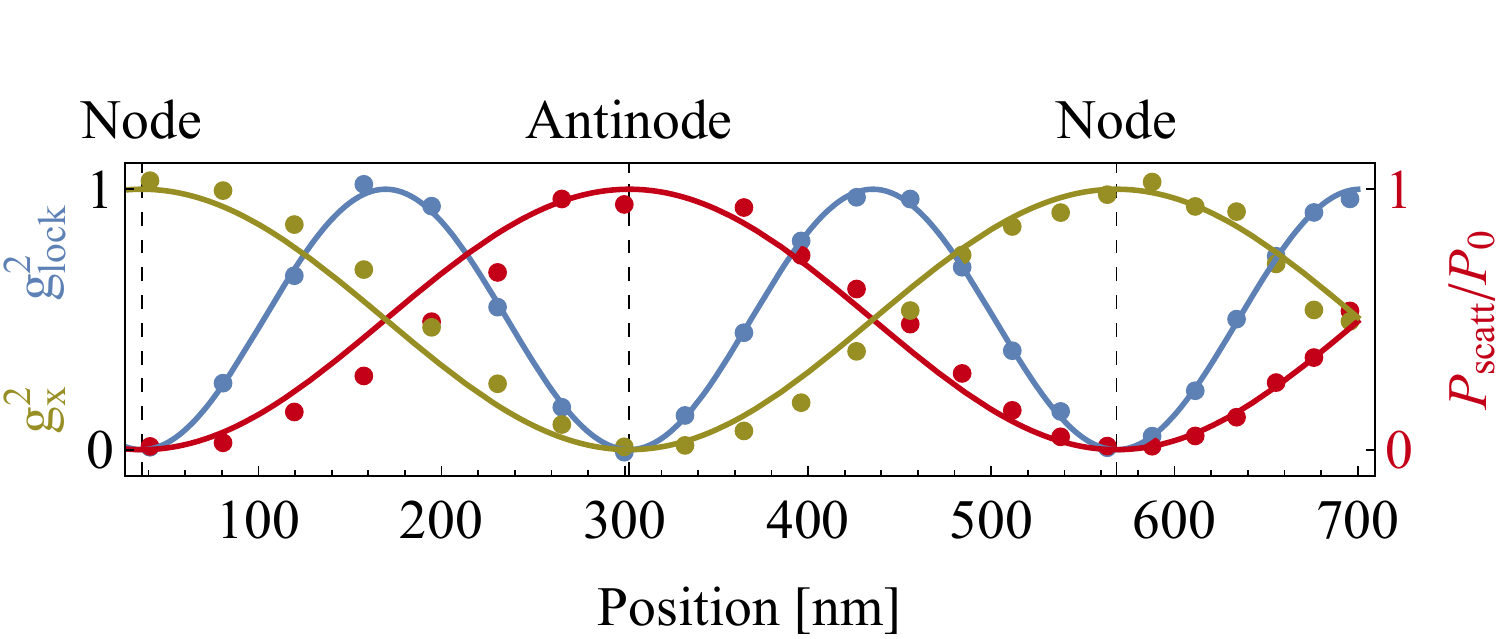}
	\caption{\label{fig:partpos} Positioning of the particle based on different detection schemes. We extract the coupling of the $x$-motion to the locking cavity mode $g_{lock}^2$ from the homodyne measurement (blue), demonstrating the standard optomechanical periodicity $g_{lock}\propto \sin(2kx_0)$. Coupling to the cavity mode populated by coherent scattering $g_x\propto \sin kx_0$ is derived from the heterodyne detection (green), where we keep the trap laser far detuned from the cavity resonance by $\Delta=2\pi\times 4$ MHz in order not to disturb the particle motion. Furthermore, the power scattered out of the cavity (red) is seen out-of-phase with $g_x$. We are able to reconstruct the nodes and antinodes of the cavity mode used for cavity cooling by coherent scattering.}
\end{figure}

In the main text, we mainly focus on the enhancement of the coherently scattered light (detector power, III) to determine the particle position $x_0$. However, the actual process involves the homodyne detection of the locking cavity mode (homodyne, II) and the heterodyne detection of the scattered photons (heterodyne, IV), which are proportional to the particle $x$-motion with $g_{lock}^2\propto \sin^2(2kx_0)$ and $\propto g_x^2\propto \sin^2(kx_0)$, respectively. This information is used to determine the cavity node and antinode of the cavity mode used for the enhancement of the coherent scattering (Fig. \ref{fig:partpos}). We show that the coupling to the locking mode governed by standard optomechanical interaction (blue) and the coupling by coherent scattering (green) follow different periodicities in particle position $x_0$, as discussed in the main text.

\subsection{Suppression of the phase noise}

The added phonon occupations and respective coupling rates to the $x$-motion in the dispersive regime and in the case of coherent scattering are:
\begin{eqnarray}
\bar{n}_{x}^{phase,disp}&=&\frac{(E_d^{disp})^2}{\kappa\left(\left(\frac{\kappa}{2}\right)^2+\Omega_x^2\right)}S_{\dot{\phi}\dot{\phi}}(\Omega_x),\hspace{1cm}\bar{n}_{x}^{phase,coh}=\frac{E_d^2\cos^2kx_0}{\kappa\left(\left(\frac{\kappa}{2}\right)^2+\Omega_x^2\right)}S_{\dot{\phi}\dot{\phi}}(\Omega_x)\nonumber\\
g_x^{disp}&=&g_0\frac{E_d^{disp}}{\sqrt{\left(\frac{\kappa}{2}\right)^2+\Omega_x^2}},\hspace{3cm}g_x=E_dkx_{zpf},
\end{eqnarray}
where $g_0=2\pi\times 0.3$~Hz is the dispersive single photon coupling of the $x$-motion of an equal-sized particle to the cavity mode \cite{Delicb2018}. Assuming that we would reach equal coupling rates $g_x^{disp}=g_x$ in the two coupling scenarios, the required cavity drive in the dispersive regime is $E_d^{disp}/2\pi\approx 4.2\times 10^{10}$~Hz. The ratio of added phonon occupations due to the phase noise heating is:
\begin{equation}
\frac{\left.\bar{n}_x^{phase,coh}\right|_{\text{node}}}{\bar{n}_x^{phase,disp}}=\frac{g_0^2\cos^2k(\lambda/4+\delta x)}{k^2x_{zpf}^2\left(\left(\frac{\kappa}{2}\right)^2+\Omega_x^2\right)},
\end{equation}
where $\delta x$ is the distance from the particle position to the cavity node. In the experiment we positioned the particle within $\delta x\approx 20$ nm and observed $50$ times less intracavity photons compared to the cavity antinode position. We estimate a decrease of the phase noise heating by a factor of $\sim 1.5\times 10^4$. More precise positioning is available, with the current nanopositioner step size of $8$~nm promising further improvement in the phase noise suppression. 

In the case of three-dimensional cavity cooling, the particle is located at the largest intensity gradient ($\cos^2kx_0=1/2$) with the measured coupling rate $g_x=2\pi\times 20$~kHz, which is the optimal position for the dispersive coupling. There, the required cavity drive in the dispersive regime would be $E_d^{disp}/2\pi=1.3\times 10^{10}$~Hz. Even in this case, the phase noise heating would be suppressed by:
\begin{equation}
\frac{\left.\bar{n}_x^{phase,coh}\right|_{\text{gradient}}}{\bar{n}_x^{phase,disp}}\approx \frac{1}{60}.
\end{equation}

In conclusion, the proximity to the intensity minimum (optimal position for the cavity cooling of the $x$-motion) results in minimal coupling of the phase noise into the cavity. Furthermore, we realize an equal coupling rate by applying a lower cavity drive in the case of coherent scattering, which additionally decreases the constraint on the phase noise.

\subsection{Quadratic cavity cooling of the $x$-motion}

%&=&-E_d\cos k(x_0+\hat{x})(\hat{a}^\dagger e^{ik\hat{z}}+\hat{a}e^{-ik\hat{z}})\nonumber\\

At the cavity antinode the interaction to the $x$-motion is intrinsically quadratic with a quadratic coupling rate:
\begin{equation}
g_{x,\text{quad}}=E_d k^2 x_{zpf}^2/2.
\end{equation}
The cooling rate is $\Gamma_{\downarrow,x}=g_{x,\text{quad}}^2\kappa/|\kappa/2+i(2\Omega_x-\Delta)|^2$, where $\kappa$ is the cavity decay rate, $\Delta$ is the trap laser detuning and $\Omega_x$ is the mechanical frequency of the $x$-motion. At pressures $p\lesssim 4$ mbar the condition $\gamma_{gas}<\Gamma_\downarrow n_{th}$ is met, such that the nonlinear damping due to quadratic cavity cooling leads to a change in phonon number distribution and to an effective cooling \cite{Nunnenkampb2010}. At pressure $p=6\times 10^{-2}~$ mbar the effective temperature of the particle motion due to the quadratic cavity cooling is:
\begin{equation}
\frac{T_{\text{quad}}^x}{T_0}=\sqrt{\frac{\gamma_{gas}}{\pi\Gamma_{\downarrow}n_{th}}}\approx 0.11,
\end{equation}
where $n_{th}=k_BT_0/(\hbar\Omega_x)$ is the thermal phonon number. In the main text, we assume a temperature model for the $x$-motion:
\begin{eqnarray}
\frac{T_{\text{eff}}^{x}(x_0)}{T_0}&=&\frac{1}{T_0}\frac{1}{\frac{\sin^2 kx_0}{T_{\text{lin}}^{x}}+\frac{\cos^2 kx_0}{T_{\text{quad}}^{x}}}
\end{eqnarray}
which is entirely parametrized by the minimum and maximum temperatures $T_{\text{lin}}^{x}$ and $T_{\text{quad}}^{x}$. The effective temperature of the $x$-motion at the cavity node is calculated from the fit of the effective damping as $T_{\text{lin}}^{x}/T_0=\gamma_{gas}/\gamma_{max}^{x}$.

\makeatletter
\renewcommand*{\@biblabel}[1]{\hfill#1.}
\makeatother

\end{widetext}


\begin{thebibliography}{48}%
	\makeatletter
	\providecommand \@ifxundefined [1]{%
		\@ifx{#1\undefined}
	}%
	\providecommand \@ifnum [1]{%
		\ifnum #1\expandafter \@firstoftwo
		\else \expandafter \@secondoftwo
		\fi
	}%
	\providecommand \@ifx [1]{%
		\ifx #1\expandafter \@firstoftwo
		\else \expandafter \@secondoftwo
		\fi
	}%
	\providecommand \natexlab [1]{#1}%
	\providecommand \enquote  [1]{``#1''}%
	\providecommand \bibnamefont  [1]{#1}%
	\providecommand \bibfnamefont [1]{#1}%
	\providecommand \citenamefont [1]{#1}%
	\providecommand \href@noop [0]{\@secondoftwo}%
	\providecommand \href [0]{\begingroup \@sanitize@url \@href}%
	\providecommand \@href[1]{\@@startlink{#1}\@@href}%
	\providecommand \@@href[1]{\endgroup#1\@@endlink}%
	\providecommand \@sanitize@url [0]{\catcode `\\12\catcode `\$12\catcode
		`\&12\catcode `\#12\catcode `\^12\catcode `\_12\catcode `\%12\relax}%
	\providecommand \@@startlink[1]{}%
	\providecommand \@@endlink[0]{}%
	\providecommand \url  [0]{\begingroup\@sanitize@url \@url }%
	\providecommand \@url [1]{\endgroup\@href {#1}{\urlprefix }}%
	\providecommand \urlprefix  [0]{URL }%
	\providecommand \Eprint [0]{\href }%
	\providecommand \doibase [0]{http://dx.doi.org/}%
	\providecommand \selectlanguage [0]{\@gobble}%
	\providecommand \bibinfo  [0]{\@secondoftwo}%
	\providecommand \bibfield  [0]{\@secondoftwo}%
	\providecommand \translation [1]{[#1]}%
	\providecommand \BibitemOpen [0]{}%
	\providecommand \bibitemStop [0]{}%
	\providecommand \bibitemNoStop [0]{.\EOS\space}%
	\providecommand \EOS [0]{\spacefactor3000\relax}%
	\providecommand \BibitemShut  [1]{\csname bibitem#1\endcsname}%
	\let\auto@bib@innerbib\@empty
	%</preamble>
	\bibitem [{\citenamefont {Phillips}(1998)}]{Phillips1998}%
	\BibitemOpen
	\bibfield  {author} {\bibinfo {author} {\bibfnamefont {W.~D.}\ \bibnamefont
			{Phillips}},\ }\href {\doibase 10.1103/RevModPhys.70.721} {\bibfield
		{journal} {\bibinfo  {journal} {Reviews of Modern Physics}\ }\textbf
		{\bibinfo {volume} {70}},\ \bibinfo {pages} {721} (\bibinfo {year}
		{1998})}\BibitemShut {NoStop}%
	\bibitem [{\citenamefont {Metcalf}\ and\ \citenamefont {van~der
			Straten}(1999)}]{Metcalf1999}%
	\BibitemOpen
	\bibfield  {author} {\bibinfo {author} {\bibfnamefont {H.~J.}\ \bibnamefont
			{Metcalf}}\ and\ \bibinfo {author} {\bibfnamefont {P.}~\bibnamefont {van~der
				Straten}},\ }\href {\doibase 10.1007/978-1-4612-1470-0} {\emph {\bibinfo
			{title} {{Laser Cooling and Trapping}}}},\ Graduate Texts in Contemporary
	Physics\ (\bibinfo  {publisher} {Springer New York},\ \bibinfo {address} {New
		York, NY},\ \bibinfo {year} {1999})\BibitemShut {NoStop}%
	\bibitem [{\citenamefont {Cohen-Tannoudji}\ and\ \citenamefont
		{Guery-Odelin}(2011)}]{Cohen-Tannoudji2011}%
	\BibitemOpen
	\bibfield  {author} {\bibinfo {author} {\bibfnamefont {C.}~\bibnamefont
			{Cohen-Tannoudji}}\ and\ \bibinfo {author} {\bibfnamefont {D.}~\bibnamefont
			{Guery-Odelin}},\ }\href@noop {} {\emph {\bibinfo {title} {{Advances in
					Atomic Physics}}}},\ \bibinfo {edition} {1st}\ ed.\ (\bibinfo  {publisher}
	{World Scientific Publishing Company},\ \bibinfo {year} {2011})\ p.\ \bibinfo
	{pages} {794}\BibitemShut {NoStop}%
	\bibitem [{\citenamefont {{Hood}}\ \emph {et~al.}(2000)\citenamefont {{Hood}},
		\citenamefont {{Lynn}}, \citenamefont {{Doherty}}, \citenamefont
		{{Parkins}},\ and\ \citenamefont {{Kimble}}}]{Hood2000}%
	\BibitemOpen
	\bibfield  {author} {\bibinfo {author} {\bibfnamefont {C.~J.}\ \bibnamefont
			{{Hood}}}, \bibinfo {author} {\bibfnamefont {T.~W.}\ \bibnamefont {{Lynn}}},
		\bibinfo {author} {\bibfnamefont {A.~C.}\ \bibnamefont {{Doherty}}}, \bibinfo
		{author} {\bibfnamefont {A.~S.}\ \bibnamefont {{Parkins}}}, \ and\ \bibinfo
		{author} {\bibfnamefont {H.~J.}\ \bibnamefont {{Kimble}}},\ }\href {\doibase
		10.1126/science.287.5457.1447} {\bibfield  {journal} {\bibinfo  {journal}
			{Science}\ }\textbf {\bibinfo {volume} {287}},\ \bibinfo {pages} {1447}
		(\bibinfo {year} {2000})}\BibitemShut {NoStop}%
	\bibitem [{\citenamefont {{Domokos}}\ \emph {et~al.}(2001)\citenamefont
		{{Domokos}}, \citenamefont {{Horak}},\ and\ \citenamefont
		{{Ritsch}}}]{Domokos2001}%
	\BibitemOpen
	\bibfield  {author} {\bibinfo {author} {\bibfnamefont {P.}~\bibnamefont
			{{Domokos}}}, \bibinfo {author} {\bibfnamefont {P.}~\bibnamefont {{Horak}}},
		\ and\ \bibinfo {author} {\bibfnamefont {H.}~\bibnamefont {{Ritsch}}},\
	}\href {\doibase 10.1088/0953-4075/34/2/306} {\bibfield  {journal} {\bibinfo
			{journal} {Journal of Physics B Atomic Molecular Physics}\ }\textbf {\bibinfo
			{volume} {34}},\ \bibinfo {pages} {187} (\bibinfo {year} {2001})}\BibitemShut
	{NoStop}%
	\bibitem [{\citenamefont {{Boozer}}\ \emph {et~al.}(2006)\citenamefont
		{{Boozer}}, \citenamefont {{Boca}}, \citenamefont {{Miller}}, \citenamefont
		{{Northup}},\ and\ \citenamefont {{Kimble}}}]{Boozer2006}%
	\BibitemOpen
	\bibfield  {author} {\bibinfo {author} {\bibfnamefont {A.~D.}\ \bibnamefont
			{{Boozer}}}, \bibinfo {author} {\bibfnamefont {A.}~\bibnamefont {{Boca}}},
		\bibinfo {author} {\bibfnamefont {R.}~\bibnamefont {{Miller}}}, \bibinfo
		{author} {\bibfnamefont {T.~E.}\ \bibnamefont {{Northup}}}, \ and\ \bibinfo
		{author} {\bibfnamefont {H.~J.}\ \bibnamefont {{Kimble}}},\ }\href {\doibase
		10.1103/PhysRevLett.97.083602} {\bibfield  {journal} {\bibinfo  {journal}
			{Physical Review Letters}\ }\textbf {\bibinfo {volume} {97}},\ \bibinfo {eid}
		{083602} (\bibinfo {year} {2006})},\ \Eprint
	{http://arxiv.org/abs/quant-ph/0606104} {quant-ph/0606104} \BibitemShut
	{NoStop}%
	\bibitem [{\citenamefont {Shuman}\ \emph {et~al.}(2010)\citenamefont {Shuman},
		\citenamefont {Barry},\ and\ \citenamefont {Demille}}]{Shuman2010}%
	\BibitemOpen
	\bibfield  {author} {\bibinfo {author} {\bibfnamefont {E.~S.}\ \bibnamefont
			{Shuman}}, \bibinfo {author} {\bibfnamefont {J.~F.}\ \bibnamefont {Barry}}, \
		and\ \bibinfo {author} {\bibfnamefont {D.}~\bibnamefont {Demille}},\ }\href
	{\doibase 10.1038/nature09443} {\bibfield  {journal} {\bibinfo  {journal}
			{Nature}\ }\textbf {\bibinfo {volume} {467}},\ \bibinfo {pages} {820}
		(\bibinfo {year} {2010})}\BibitemShut {NoStop}%
	\bibitem [{\citenamefont {{Hummon}}\ \emph {et~al.}(2013)\citenamefont
		{{Hummon}}, \citenamefont {{Yeo}}, \citenamefont {{Stuhl}}, \citenamefont
		{{Collopy}}, \citenamefont {{Xia}},\ and\ \citenamefont {{Ye}}}]{Hummon2013}%
	\BibitemOpen
	\bibfield  {author} {\bibinfo {author} {\bibfnamefont {M.~T.}\ \bibnamefont
			{{Hummon}}}, \bibinfo {author} {\bibfnamefont {M.}~\bibnamefont {{Yeo}}},
		\bibinfo {author} {\bibfnamefont {B.~K.}\ \bibnamefont {{Stuhl}}}, \bibinfo
		{author} {\bibfnamefont {A.~L.}\ \bibnamefont {{Collopy}}}, \bibinfo {author}
		{\bibfnamefont {Y.}~\bibnamefont {{Xia}}}, \ and\ \bibinfo {author}
		{\bibfnamefont {J.}~\bibnamefont {{Ye}}},\ }\href {\doibase
		10.1103/PhysRevLett.110.143001} {\bibfield  {journal} {\bibinfo  {journal}
			{Physical Review Letters}\ }\textbf {\bibinfo {volume} {110}},\ \bibinfo
		{eid} {143001} (\bibinfo {year} {2013})}\BibitemShut {NoStop}%
	\bibitem [{\citenamefont {Lim}\ \emph {et~al.}(2018)\citenamefont {Lim},
		\citenamefont {Almond}, \citenamefont {Trigatzis}, \citenamefont {Devlin},
		\citenamefont {Fitch}, \citenamefont {Sauer}, \citenamefont {Tarbutt},\ and\
		\citenamefont {Hinds}}]{Lim2018}%
	\BibitemOpen
	\bibfield  {author} {\bibinfo {author} {\bibfnamefont {J.}~\bibnamefont
			{Lim}}, \bibinfo {author} {\bibfnamefont {J.~R.}\ \bibnamefont {Almond}},
		\bibinfo {author} {\bibfnamefont {M.~A.}\ \bibnamefont {Trigatzis}}, \bibinfo
		{author} {\bibfnamefont {J.~A.}\ \bibnamefont {Devlin}}, \bibinfo {author}
		{\bibfnamefont {N.~J.}\ \bibnamefont {Fitch}}, \bibinfo {author}
		{\bibfnamefont {B.~E.}\ \bibnamefont {Sauer}}, \bibinfo {author}
		{\bibfnamefont {M.~R.}\ \bibnamefont {Tarbutt}}, \ and\ \bibinfo {author}
		{\bibfnamefont {E.~A.}\ \bibnamefont {Hinds}},\ }\href {\doibase
		10.1103/PhysRevLett.120.123201} {\bibfield  {journal} {\bibinfo  {journal}
			{Physical Review Letters}\ }\textbf {\bibinfo {volume} {120}},\ \bibinfo
		{pages} {123201} (\bibinfo {year} {2018})}\BibitemShut {NoStop}%
	\bibitem [{\citenamefont {Anderegg}\ \emph {et~al.}(2018)\citenamefont
		{Anderegg}, \citenamefont {Augenbraun}, \citenamefont {Bao}, \citenamefont
		{Burchesky}, \citenamefont {Cheuk}, \citenamefont {Ketterle},\ and\
		\citenamefont {Doyle}}]{Anderegg2018}%
	\BibitemOpen
	\bibfield  {author} {\bibinfo {author} {\bibfnamefont {L.}~\bibnamefont
			{Anderegg}}, \bibinfo {author} {\bibfnamefont {B.~L.}\ \bibnamefont
			{Augenbraun}}, \bibinfo {author} {\bibfnamefont {Y.}~\bibnamefont {Bao}},
		\bibinfo {author} {\bibfnamefont {S.}~\bibnamefont {Burchesky}}, \bibinfo
		{author} {\bibfnamefont {L.~W.}\ \bibnamefont {Cheuk}}, \bibinfo {author}
		{\bibfnamefont {W.}~\bibnamefont {Ketterle}}, \ and\ \bibinfo {author}
		{\bibfnamefont {J.~M.}\ \bibnamefont {Doyle}},\ }\href {\doibase
		10.1038/s41567-018-0191-z} {\bibfield  {journal} {\bibinfo  {journal} {Nature
				Physics}\ }\textbf {\bibinfo {volume} {14}},\ \bibinfo {pages} {890}
		(\bibinfo {year} {2018})}\BibitemShut {NoStop}%
	\bibitem [{\citenamefont {McCarron}(2018)}]{McCarron2018}%
	\BibitemOpen
	\bibfield  {author} {\bibinfo {author} {\bibfnamefont {D.}~\bibnamefont
			{McCarron}},\ }\href {\doibase 10.1088/1361-6455/aadfba} {\bibfield
		{journal} {\bibinfo  {journal} {Journal of Physics B: Atomic, Molecular and
				Optical Physics}\ }\textbf {\bibinfo {volume} {51}},\ \bibinfo {pages}
		{212001} (\bibinfo {year} {2018})}\BibitemShut {NoStop}%
	\bibitem [{\citenamefont {{Vuleti{\'c}}}\ and\ \citenamefont
		{{Chu}}(2000)}]{Vuletic2000}%
	\BibitemOpen
	\bibfield  {author} {\bibinfo {author} {\bibfnamefont {V.}~\bibnamefont
			{{Vuleti{\'c}}}}\ and\ \bibinfo {author} {\bibfnamefont {S.}~\bibnamefont
			{{Chu}}},\ }\href {\doibase 10.1103/PhysRevLett.84.3787} {\bibfield
		{journal} {\bibinfo  {journal} {Physical Review Letters}\ }\textbf {\bibinfo
			{volume} {84}},\ \bibinfo {pages} {3787} (\bibinfo {year}
		{2000})}\BibitemShut {NoStop}%
	\bibitem [{\citenamefont {{Horak}}\ \emph {et~al.}(1997)\citenamefont
		{{Horak}}, \citenamefont {{Hechenblaikner}}, \citenamefont {{Gheri}},
		\citenamefont {{Stecher}},\ and\ \citenamefont {{Ritsch}}}]{Horak1997}%
	\BibitemOpen
	\bibfield  {author} {\bibinfo {author} {\bibfnamefont {P.}~\bibnamefont
			{{Horak}}}, \bibinfo {author} {\bibfnamefont {G.}~\bibnamefont
			{{Hechenblaikner}}}, \bibinfo {author} {\bibfnamefont {K.~M.}\ \bibnamefont
			{{Gheri}}}, \bibinfo {author} {\bibfnamefont {H.}~\bibnamefont {{Stecher}}},
		\ and\ \bibinfo {author} {\bibfnamefont {H.}~\bibnamefont {{Ritsch}}},\
	}\href {\doibase 10.1103/PhysRevLett.79.4974} {\bibfield  {journal} {\bibinfo
			{journal} {Physical Review Letters}\ }\textbf {\bibinfo {volume} {79}},\
		\bibinfo {pages} {4974} (\bibinfo {year} {1997})}\BibitemShut {NoStop}%
	\bibitem [{\citenamefont {{Chan}}\ \emph {et~al.}(2003)\citenamefont {{Chan}},
		\citenamefont {{Black}},\ and\ \citenamefont {{Vuleti{\'c}}}}]{Chan2003}%
	\BibitemOpen
	\bibfield  {author} {\bibinfo {author} {\bibfnamefont {H.~W.}\ \bibnamefont
			{{Chan}}}, \bibinfo {author} {\bibfnamefont {A.~T.}\ \bibnamefont {{Black}}},
		\ and\ \bibinfo {author} {\bibfnamefont {V.}~\bibnamefont {{Vuleti{\'c}}}},\
	}\href {\doibase 10.1103/PhysRevLett.90.063003} {\bibfield  {journal}
		{\bibinfo  {journal} {Physical Review Letters}\ }\textbf {\bibinfo {volume}
			{90}},\ \bibinfo {eid} {063003} (\bibinfo {year} {2003})},\ \Eprint
	{http://arxiv.org/abs/quant-ph/0208100} {quant-ph/0208100} \BibitemShut
	{NoStop}%
	\bibitem [{\citenamefont {{Maunz}}\ \emph {et~al.}(2004)\citenamefont
		{{Maunz}}, \citenamefont {{Puppe}}, \citenamefont {{Schuster}}, \citenamefont
		{{Syassen}}, \citenamefont {{Pinkse}},\ and\ \citenamefont
		{{Rempe}}}]{Maunz2004}%
	\BibitemOpen
	\bibfield  {author} {\bibinfo {author} {\bibfnamefont {P.}~\bibnamefont
			{{Maunz}}}, \bibinfo {author} {\bibfnamefont {T.}~\bibnamefont {{Puppe}}},
		\bibinfo {author} {\bibfnamefont {I.}~\bibnamefont {{Schuster}}}, \bibinfo
		{author} {\bibfnamefont {N.}~\bibnamefont {{Syassen}}}, \bibinfo {author}
		{\bibfnamefont {P.~W.~H.}\ \bibnamefont {{Pinkse}}}, \ and\ \bibinfo {author}
		{\bibfnamefont {G.}~\bibnamefont {{Rempe}}},\ }\href {\doibase
		10.1038/nature02387} {\bibfield  {journal} {\bibinfo  {journal} {\nat}\
		}\textbf {\bibinfo {volume} {428}},\ \bibinfo {pages} {50} (\bibinfo {year}
		{2004})}\BibitemShut {NoStop}%
	\bibitem [{\citenamefont {{Nu{\ss}mann}}\ \emph {et~al.}(2005)\citenamefont
		{{Nu{\ss}mann}}, \citenamefont {{Murr}}, \citenamefont {{Hijlkema}},
		\citenamefont {{Weber}}, \citenamefont {{Kuhn}},\ and\ \citenamefont
		{{Rempe}}}]{Nussmann2005}%
	\BibitemOpen
	\bibfield  {author} {\bibinfo {author} {\bibfnamefont {S.}~\bibnamefont
			{{Nu{\ss}mann}}}, \bibinfo {author} {\bibfnamefont {K.}~\bibnamefont
			{{Murr}}}, \bibinfo {author} {\bibfnamefont {M.}~\bibnamefont {{Hijlkema}}},
		\bibinfo {author} {\bibfnamefont {B.}~\bibnamefont {{Weber}}}, \bibinfo
		{author} {\bibfnamefont {A.}~\bibnamefont {{Kuhn}}}, \ and\ \bibinfo {author}
		{\bibfnamefont {G.}~\bibnamefont {{Rempe}}},\ }\href {\doibase
		10.1038/nphys120} {\bibfield  {journal} {\bibinfo  {journal} {Nature
				Physics}\ }\textbf {\bibinfo {volume} {1}},\ \bibinfo {pages} {122} (\bibinfo
		{year} {2005})}\BibitemShut {NoStop}%
	\bibitem [{\citenamefont {{Fortier}}\ \emph {et~al.}(2007)\citenamefont
		{{Fortier}}, \citenamefont {{Kim}}, \citenamefont {{Gibbons}}, \citenamefont
		{{Ahmadi}},\ and\ \citenamefont {{Chapman}}}]{Fortier2007}%
	\BibitemOpen
	\bibfield  {author} {\bibinfo {author} {\bibfnamefont {K.~M.}\ \bibnamefont
			{{Fortier}}}, \bibinfo {author} {\bibfnamefont {S.~Y.}\ \bibnamefont
			{{Kim}}}, \bibinfo {author} {\bibfnamefont {M.~J.}\ \bibnamefont
			{{Gibbons}}}, \bibinfo {author} {\bibfnamefont {P.}~\bibnamefont {{Ahmadi}}},
		\ and\ \bibinfo {author} {\bibfnamefont {M.~S.}\ \bibnamefont {{Chapman}}},\
	}\href {\doibase 10.1103/PhysRevLett.98.233601} {\bibfield  {journal}
		{\bibinfo  {journal} {Physical Review Letters}\ }\textbf {\bibinfo {volume}
			{98}},\ \bibinfo {eid} {233601} (\bibinfo {year} {2007})}\BibitemShut
	{NoStop}%
	\bibitem [{\citenamefont {{Gigan}}\ \emph {et~al.}(2006)\citenamefont
		{{Gigan}}, \citenamefont {{B{\"o}hm}}, \citenamefont {{Paternostro}},
		\citenamefont {{Blaser}}, \citenamefont {{Langer}}, \citenamefont
		{{Hertzberg}}, \citenamefont {{Schwab}}, \citenamefont {{B{\"a}uerle}},
		\citenamefont {{Aspelmeyer}},\ and\ \citenamefont {{Zeilinger}}}]{Gigan2006}%
	\BibitemOpen
	\bibfield  {author} {\bibinfo {author} {\bibfnamefont {S.}~\bibnamefont
			{{Gigan}}}, \bibinfo {author} {\bibfnamefont {H.~R.}\ \bibnamefont
			{{B{\"o}hm}}}, \bibinfo {author} {\bibfnamefont {M.}~\bibnamefont
			{{Paternostro}}}, \bibinfo {author} {\bibfnamefont {F.}~\bibnamefont
			{{Blaser}}}, \bibinfo {author} {\bibfnamefont {G.}~\bibnamefont {{Langer}}},
		\bibinfo {author} {\bibfnamefont {J.~B.}\ \bibnamefont {{Hertzberg}}},
		\bibinfo {author} {\bibfnamefont {K.~C.}\ \bibnamefont {{Schwab}}}, \bibinfo
		{author} {\bibfnamefont {D.}~\bibnamefont {{B{\"a}uerle}}}, \bibinfo {author}
		{\bibfnamefont {M.}~\bibnamefont {{Aspelmeyer}}}, \ and\ \bibinfo {author}
		{\bibfnamefont {A.}~\bibnamefont {{Zeilinger}}},\ }\href {\doibase
		10.1038/nature05273} {\bibfield  {journal} {\bibinfo  {journal} {\nat}\
		}\textbf {\bibinfo {volume} {444}},\ \bibinfo {pages} {67} (\bibinfo {year}
		{2006})}\BibitemShut {NoStop}%
	\bibitem [{\citenamefont {Arcizet}\ \emph {et~al.}(2006)\citenamefont
		{Arcizet}, \citenamefont {Cohadon}, \citenamefont {Briant}, \citenamefont
		{Pinard},\ and\ \citenamefont {Heidmann}}]{Arcizet2006a}%
	\BibitemOpen
	\bibfield  {author} {\bibinfo {author} {\bibfnamefont {O.}~\bibnamefont
			{Arcizet}}, \bibinfo {author} {\bibfnamefont {P.-F.~F.}\ \bibnamefont
			{Cohadon}}, \bibinfo {author} {\bibfnamefont {T.}~\bibnamefont {Briant}},
		\bibinfo {author} {\bibfnamefont {M.}~\bibnamefont {Pinard}}, \ and\ \bibinfo
		{author} {\bibfnamefont {A.}~\bibnamefont {Heidmann}},\ }\href@noop {}
	{\bibfield  {journal} {\bibinfo  {journal} {Nature}\ }\textbf {\bibinfo
			{volume} {444}},\ \bibinfo {pages} {71} (\bibinfo {year} {2006})}\BibitemShut
	{NoStop}%
	\bibitem [{\citenamefont {{Schliesser}}\ \emph {et~al.}(2006)\citenamefont
		{{Schliesser}}, \citenamefont {{Del'Haye}}, \citenamefont {{Nooshi}},
		\citenamefont {{Vahala}},\ and\ \citenamefont
		{{Kippenberg}}}]{Schliesser2006}%
	\BibitemOpen
	\bibfield  {author} {\bibinfo {author} {\bibfnamefont {A.}~\bibnamefont
			{{Schliesser}}}, \bibinfo {author} {\bibfnamefont {P.}~\bibnamefont
			{{Del'Haye}}}, \bibinfo {author} {\bibfnamefont {N.}~\bibnamefont
			{{Nooshi}}}, \bibinfo {author} {\bibfnamefont {K.~J.}\ \bibnamefont
			{{Vahala}}}, \ and\ \bibinfo {author} {\bibfnamefont {T.~J.}\ \bibnamefont
			{{Kippenberg}}},\ }\href {\doibase 10.1103/PhysRevLett.97.243905} {\bibfield
		{journal} {\bibinfo  {journal} {Physical Review Letters}\ }\textbf {\bibinfo
			{volume} {97}},\ \bibinfo {eid} {243905} (\bibinfo {year}
		{2006})}\BibitemShut {NoStop}%
	\bibitem [{\citenamefont {Thompson}\ \emph {et~al.}(2008)\citenamefont
		{Thompson}, \citenamefont {Zwickl}, \citenamefont {Jayich}, \citenamefont
		{Marquardt}, \citenamefont {Girvin},\ and\ \citenamefont
		{Harris}}]{Thompson2008}%
	\BibitemOpen
	\bibfield  {author} {\bibinfo {author} {\bibfnamefont {J.~D.}\ \bibnamefont
			{Thompson}}, \bibinfo {author} {\bibfnamefont {B.~M.}\ \bibnamefont
			{Zwickl}}, \bibinfo {author} {\bibfnamefont {A.~M.}\ \bibnamefont {Jayich}},
		\bibinfo {author} {\bibfnamefont {F.}~\bibnamefont {Marquardt}}, \bibinfo
		{author} {\bibfnamefont {S.~M.}\ \bibnamefont {Girvin}}, \ and\ \bibinfo
		{author} {\bibfnamefont {J.~G.~E.}\ \bibnamefont {Harris}},\ }\href@noop {}
	{\bibfield  {journal} {\bibinfo  {journal} {Nature}\ }\textbf {\bibinfo
			{volume} {452}},\ \bibinfo {pages} {72} (\bibinfo {year} {2008})}\BibitemShut
	{NoStop}%
	\bibitem [{\citenamefont {Teufel}\ \emph {et~al.}(2011)\citenamefont {Teufel},
		\citenamefont {Donner}, \citenamefont {Li}, \citenamefont {Harlow},
		\citenamefont {Allman}, \citenamefont {Cicak}, \citenamefont {Sirois},
		\citenamefont {Whittaker}, \citenamefont {Lehnert},\ and\ \citenamefont
		{Simmonds}}]{Teufel2011}%
	\BibitemOpen
	\bibfield  {author} {\bibinfo {author} {\bibfnamefont {J.~D.}\ \bibnamefont
			{Teufel}}, \bibinfo {author} {\bibfnamefont {T.}~\bibnamefont {Donner}},
		\bibinfo {author} {\bibfnamefont {D.}~\bibnamefont {Li}}, \bibinfo {author}
		{\bibfnamefont {J.~W.}\ \bibnamefont {Harlow}}, \bibinfo {author}
		{\bibfnamefont {M.~S.}\ \bibnamefont {Allman}}, \bibinfo {author}
		{\bibfnamefont {K.}~\bibnamefont {Cicak}}, \bibinfo {author} {\bibfnamefont
			{a.~J.}\ \bibnamefont {Sirois}}, \bibinfo {author} {\bibfnamefont {J.~D.}\
			\bibnamefont {Whittaker}}, \bibinfo {author} {\bibfnamefont {K.~W.}\
			\bibnamefont {Lehnert}}, \ and\ \bibinfo {author} {\bibfnamefont {R.~W.}\
			\bibnamefont {Simmonds}},\ }\href {\doibase 10.1038/nature10261} {\bibfield
		{journal} {\bibinfo  {journal} {Nature}\ }\textbf {\bibinfo {volume} {475}},\
		\bibinfo {pages} {359} (\bibinfo {year} {2011})}\BibitemShut {NoStop}%
	\bibitem [{\citenamefont {Chan}\ \emph {et~al.}(2011)\citenamefont {Chan},
		\citenamefont {Alegre}, \citenamefont {Safavi-Naeini}, \citenamefont {Hill},
		\citenamefont {Krause}, \citenamefont {Gr{\"{o}}blacher}, \citenamefont
		{Aspelmeyer},\ and\ \citenamefont {Painter}}]{Chan2011c}%
	\BibitemOpen
	\bibfield  {author} {\bibinfo {author} {\bibfnamefont {J.}~\bibnamefont
			{Chan}}, \bibinfo {author} {\bibfnamefont {T.~P.~M.}\ \bibnamefont {Alegre}},
		\bibinfo {author} {\bibfnamefont {A.~H.}\ \bibnamefont {Safavi-Naeini}},
		\bibinfo {author} {\bibfnamefont {J.~T.}\ \bibnamefont {Hill}}, \bibinfo
		{author} {\bibfnamefont {A.}~\bibnamefont {Krause}}, \bibinfo {author}
		{\bibfnamefont {S.}~\bibnamefont {Gr{\"{o}}blacher}}, \bibinfo {author}
		{\bibfnamefont {M.}~\bibnamefont {Aspelmeyer}}, \ and\ \bibinfo {author}
		{\bibfnamefont {O.}~\bibnamefont {Painter}},\ }\href {\doibase
		10.1038/nature10461} {\bibfield  {journal} {\bibinfo  {journal} {Nature}\
		}\textbf {\bibinfo {volume} {478}},\ \bibinfo {pages} {89} (\bibinfo {year}
		{2011})}\BibitemShut {NoStop}%
	\bibitem [{\citenamefont {{Aspelmeyer}}\ \emph {et~al.}(2014)\citenamefont
		{{Aspelmeyer}}, \citenamefont {{Kippenberg}},\ and\ \citenamefont
		{{Marquardt}}}]{RMP2014}%
	\BibitemOpen
	\bibfield  {author} {\bibinfo {author} {\bibfnamefont {M.}~\bibnamefont
			{{Aspelmeyer}}}, \bibinfo {author} {\bibfnamefont {T.~J.}\ \bibnamefont
			{{Kippenberg}}}, \ and\ \bibinfo {author} {\bibfnamefont {F.}~\bibnamefont
			{{Marquardt}}},\ }\href {\doibase 10.1103/RevModPhys.86.1391} {\bibfield
		{journal} {\bibinfo  {journal} {Reviews of Modern Physics}\ }\textbf
		{\bibinfo {volume} {86}},\ \bibinfo {pages} {1391} (\bibinfo {year}
		{2014})}\BibitemShut {NoStop}%
	\bibitem [{\citenamefont {{Kiesel}}\ \emph {et~al.}(2013)\citenamefont
		{{Kiesel}}, \citenamefont {{Blaser}}, \citenamefont {{Deli{\'c}}},
		\citenamefont {{Grass}}, \citenamefont {{Kaltenbaek}},\ and\ \citenamefont
		{{Aspelmeyer}}}]{Kiesel2013}%
	\BibitemOpen
	\bibfield  {author} {\bibinfo {author} {\bibfnamefont {N.}~\bibnamefont
			{{Kiesel}}}, \bibinfo {author} {\bibfnamefont {F.}~\bibnamefont {{Blaser}}},
		\bibinfo {author} {\bibfnamefont {U.}~\bibnamefont {{Deli{\'c}}}}, \bibinfo
		{author} {\bibfnamefont {D.}~\bibnamefont {{Grass}}}, \bibinfo {author}
		{\bibfnamefont {R.}~\bibnamefont {{Kaltenbaek}}}, \ and\ \bibinfo {author}
		{\bibfnamefont {M.}~\bibnamefont {{Aspelmeyer}}},\ }\href {\doibase
		10.1073/pnas.1309167110} {\bibfield  {journal} {\bibinfo  {journal}
			{Proceedings of the National Academy of Science}\ }\textbf {\bibinfo {volume}
			{110}},\ \bibinfo {pages} {14180} (\bibinfo {year} {2013})}\BibitemShut
	{NoStop}%
	\bibitem [{\citenamefont {{Asenbaum}}\ \emph {et~al.}(2013)\citenamefont
		{{Asenbaum}}, \citenamefont {{Kuhn}}, \citenamefont {{Nimmrichter}},
		\citenamefont {{Sezer}},\ and\ \citenamefont {{Arndt}}}]{Asenbaum2013}%
	\BibitemOpen
	\bibfield  {author} {\bibinfo {author} {\bibfnamefont {P.}~\bibnamefont
			{{Asenbaum}}}, \bibinfo {author} {\bibfnamefont {S.}~\bibnamefont {{Kuhn}}},
		\bibinfo {author} {\bibfnamefont {S.}~\bibnamefont {{Nimmrichter}}}, \bibinfo
		{author} {\bibfnamefont {U.}~\bibnamefont {{Sezer}}}, \ and\ \bibinfo
		{author} {\bibfnamefont {M.}~\bibnamefont {{Arndt}}},\ }\href {\doibase
		10.1038/ncomms3743} {\bibfield  {journal} {\bibinfo  {journal} {Nature
				Communications}\ }\textbf {\bibinfo {volume} {4}},\ \bibinfo {eid} {2743}
		(\bibinfo {year} {2013})}\BibitemShut {NoStop}%
	\bibitem [{\citenamefont {{Millen}}\ \emph {et~al.}(2015)\citenamefont
		{{Millen}}, \citenamefont {{Fonseca}}, \citenamefont {{Mavrogordatos}},
		\citenamefont {{Monteiro}},\ and\ \citenamefont {{Barker}}}]{Millen2015}%
	\BibitemOpen
	\bibfield  {author} {\bibinfo {author} {\bibfnamefont {J.}~\bibnamefont
			{{Millen}}}, \bibinfo {author} {\bibfnamefont {P.~Z.~G.}\ \bibnamefont
			{{Fonseca}}}, \bibinfo {author} {\bibfnamefont {T.}~\bibnamefont
			{{Mavrogordatos}}}, \bibinfo {author} {\bibfnamefont {T.~S.}\ \bibnamefont
			{{Monteiro}}}, \ and\ \bibinfo {author} {\bibfnamefont {P.~F.}\ \bibnamefont
			{{Barker}}},\ }\href {\doibase 10.1103/PhysRevLett.114.123602} {\bibfield
		{journal} {\bibinfo  {journal} {Physical Review Letters}\ }\textbf {\bibinfo
			{volume} {114}},\ \bibinfo {eid} {123602} (\bibinfo {year}
		{2015})}\BibitemShut {NoStop}%
	\bibitem [{\citenamefont {{Fonseca}}\ \emph {et~al.}(2016)\citenamefont
		{{Fonseca}}, \citenamefont {{Aranas}}, \citenamefont {{Millen}},
		\citenamefont {{Monteiro}},\ and\ \citenamefont {{Barker}}}]{Fonseca2016}%
	\BibitemOpen
	\bibfield  {author} {\bibinfo {author} {\bibfnamefont {P.~Z.~G.}\
			\bibnamefont {{Fonseca}}}, \bibinfo {author} {\bibfnamefont {E.~B.}\
			\bibnamefont {{Aranas}}}, \bibinfo {author} {\bibfnamefont {J.}~\bibnamefont
			{{Millen}}}, \bibinfo {author} {\bibfnamefont {T.~S.}\ \bibnamefont
			{{Monteiro}}}, \ and\ \bibinfo {author} {\bibfnamefont {P.~F.}\ \bibnamefont
			{{Barker}}},\ }\href {\doibase 10.1103/PhysRevLett.117.173602} {\bibfield
		{journal} {\bibinfo  {journal} {Physical Review Letters}\ }\textbf {\bibinfo
			{volume} {117}},\ \bibinfo {eid} {173602} (\bibinfo {year}
		{2016})}\BibitemShut {NoStop}%
	\bibitem [{\citenamefont {{Deli{\'c}}}\ \emph {et~al.}()\citenamefont
		{{Deli{\'c}}}, \citenamefont {{Grass}}, \citenamefont {{Reisenbauer}},
		\citenamefont {{Kiesel}},\ and\ \citenamefont {{Aspelmeyer}}}]{Delic2018}%
	\BibitemOpen
	\bibfield  {author} {\bibinfo {author} {\bibfnamefont {U.}~\bibnamefont
			{{Deli{\'c}}}}, \bibinfo {author} {\bibfnamefont {D.}~\bibnamefont
			{{Grass}}}, \bibinfo {author} {\bibfnamefont {M.}~\bibnamefont
			{{Reisenbauer}}}, \bibinfo {author} {\bibfnamefont {N.}~\bibnamefont
			{{Kiesel}}}, \ and\ \bibinfo {author} {\bibfnamefont {M.}~\bibnamefont
			{{Aspelmeyer}}},\ }\href@noop {} {\bibinfo  {journal} {(unpublished)}\
	}\BibitemShut {NoStop}%
	\bibitem [{\citenamefont {{Rabl}}\ \emph {et~al.}(2009)\citenamefont {{Rabl}},
		\citenamefont {{Genes}}, \citenamefont {{Hammerer}},\ and\ \citenamefont
		{{Aspelmeyer}}}]{Rabl2009}%
	\BibitemOpen
	\bibfield  {journal} {  }\bibfield  {author} {\bibinfo {author} {\bibfnamefont
			{P.}~\bibnamefont {{Rabl}}}, \bibinfo {author} {\bibfnamefont
			{C.}~\bibnamefont {{Genes}}}, \bibinfo {author} {\bibfnamefont
			{K.}~\bibnamefont {{Hammerer}}}, \ and\ \bibinfo {author} {\bibfnamefont
			{M.}~\bibnamefont {{Aspelmeyer}}},\ }\href {\doibase
		10.1103/PhysRevA.80.063819} {\bibfield  {journal} {\bibinfo  {journal}
			{\pra}\ }\textbf {\bibinfo {volume} {80}},\ \bibinfo {eid} {063819} (\bibinfo
		{year} {2009})}\BibitemShut {NoStop}%
	\bibitem [{\citenamefont {{Jayich}}\ \emph {et~al.}(2012)\citenamefont
		{{Jayich}}, \citenamefont {{Sankey}}, \citenamefont {{B{\o}rkje}},
		\citenamefont {{Lee}}, \citenamefont {{Yang}}, \citenamefont {{Underwood}},
		\citenamefont {{Childress}}, \citenamefont {{Petrenko}}, \citenamefont
		{{Girvin}},\ and\ \citenamefont {{Harris}}}]{Jayich2012}%
	\BibitemOpen
	\bibfield  {author} {\bibinfo {author} {\bibfnamefont {A.~M.}\ \bibnamefont
			{{Jayich}}}, \bibinfo {author} {\bibfnamefont {J.~C.}\ \bibnamefont
			{{Sankey}}}, \bibinfo {author} {\bibfnamefont {K.}~\bibnamefont
			{{B{\o}rkje}}}, \bibinfo {author} {\bibfnamefont {D.}~\bibnamefont {{Lee}}},
		\bibinfo {author} {\bibfnamefont {C.}~\bibnamefont {{Yang}}}, \bibinfo
		{author} {\bibfnamefont {M.}~\bibnamefont {{Underwood}}}, \bibinfo {author}
		{\bibfnamefont {L.}~\bibnamefont {{Childress}}}, \bibinfo {author}
		{\bibfnamefont {A.}~\bibnamefont {{Petrenko}}}, \bibinfo {author}
		{\bibfnamefont {S.~M.}\ \bibnamefont {{Girvin}}}, \ and\ \bibinfo {author}
		{\bibfnamefont {J.~G.~E.}\ \bibnamefont {{Harris}}},\ }\href {\doibase
		10.1088/1367-2630/14/11/115018} {\bibfield  {journal} {\bibinfo  {journal}
			{New Journal of Physics}\ }\textbf {\bibinfo {volume} {14}},\ \bibinfo {eid}
		{115018} (\bibinfo {year} {2012})}\BibitemShut {NoStop}%
	\bibitem [{\citenamefont {{Safavi-Naeini}}\ \emph {et~al.}(2013)\citenamefont
		{{Safavi-Naeini}}, \citenamefont {{Chan}}, \citenamefont {{Hill}},
		\citenamefont {{Gr{\"o}blacher}}, \citenamefont {{Miao}}, \citenamefont
		{{Chen}}, \citenamefont {{Aspelmeyer}},\ and\ \citenamefont
		{{Painter}}}]{Safavi-Naeini2013}%
	\BibitemOpen
	\bibfield  {author} {\bibinfo {author} {\bibfnamefont {A.~H.}\ \bibnamefont
			{{Safavi-Naeini}}}, \bibinfo {author} {\bibfnamefont {J.}~\bibnamefont
			{{Chan}}}, \bibinfo {author} {\bibfnamefont {J.~T.}\ \bibnamefont {{Hill}}},
		\bibinfo {author} {\bibfnamefont {S.}~\bibnamefont {{Gr{\"o}blacher}}},
		\bibinfo {author} {\bibfnamefont {H.}~\bibnamefont {{Miao}}}, \bibinfo
		{author} {\bibfnamefont {Y.}~\bibnamefont {{Chen}}}, \bibinfo {author}
		{\bibfnamefont {M.}~\bibnamefont {{Aspelmeyer}}}, \ and\ \bibinfo {author}
		{\bibfnamefont {O.}~\bibnamefont {{Painter}}},\ }\href {\doibase
		10.1088/1367-2630/15/3/035007} {\bibfield  {journal} {\bibinfo  {journal}
			{New Journal of Physics}\ }\textbf {\bibinfo {volume} {15}},\ \bibinfo {eid}
		{035007} (\bibinfo {year} {2013})}\BibitemShut {NoStop}%
	\bibitem [{SI()}]{SI}%
	\BibitemOpen
	\href@noop {} {}\bibinfo {note} {See Supplemental Material at [URL will be
		inserted by publisher] for the details about the theory, positioning of the
		particle and data evaluation.}\BibitemShut {Stop}%
	\bibitem [{\citenamefont {{Vuleti{\'c}}}\ \emph {et~al.}(2001)\citenamefont
		{{Vuleti{\'c}}}, \citenamefont {{Chan}},\ and\ \citenamefont
		{{Black}}}]{Vuletic2001}%
	\BibitemOpen
	\bibfield  {author} {\bibinfo {author} {\bibfnamefont {V.}~\bibnamefont
			{{Vuleti{\'c}}}}, \bibinfo {author} {\bibfnamefont {H.~W.}\ \bibnamefont
			{{Chan}}}, \ and\ \bibinfo {author} {\bibfnamefont {A.~T.}\ \bibnamefont
			{{Black}}},\ }\href {\doibase 10.1103/PhysRevA.64.033405} {\bibfield
		{journal} {\bibinfo  {journal} {\pra}\ }\textbf {\bibinfo {volume} {64}},\
		\bibinfo {eid} {033405} (\bibinfo {year} {2001})}\BibitemShut {NoStop}%
	\bibitem [{\citenamefont {{Hosseini}}\ \emph {et~al.}(2017)\citenamefont
		{{Hosseini}}, \citenamefont {{Duan}}, \citenamefont {{Beck}}, \citenamefont
		{{Chen}},\ and\ \citenamefont {{Vuleti{\'c}}}}]{Hosseini2017}%
	\BibitemOpen
	\bibfield  {author} {\bibinfo {author} {\bibfnamefont {M.}~\bibnamefont
			{{Hosseini}}}, \bibinfo {author} {\bibfnamefont {Y.}~\bibnamefont {{Duan}}},
		\bibinfo {author} {\bibfnamefont {K.~M.}\ \bibnamefont {{Beck}}}, \bibinfo
		{author} {\bibfnamefont {Y.-T.}\ \bibnamefont {{Chen}}}, \ and\ \bibinfo
		{author} {\bibfnamefont {V.}~\bibnamefont {{Vuleti{\'c}}}},\ }\href {\doibase
		10.1103/PhysRevLett.118.183601} {\bibfield  {journal} {\bibinfo  {journal}
			{Physical Review Letters}\ }\textbf {\bibinfo {volume} {118}},\ \bibinfo
		{eid} {183601} (\bibinfo {year} {2017})}\BibitemShut {NoStop}%
	\bibitem [{\citenamefont {{Leibrandt}}\ \emph {et~al.}(2009)\citenamefont
		{{Leibrandt}}, \citenamefont {{Labaziewicz}}, \citenamefont {{Vuleti{\'c}}},\
		and\ \citenamefont {{Chuang}}}]{Leibrandt2009}%
	\BibitemOpen
	\bibfield  {author} {\bibinfo {author} {\bibfnamefont {D.~R.}\ \bibnamefont
			{{Leibrandt}}}, \bibinfo {author} {\bibfnamefont {J.}~\bibnamefont
			{{Labaziewicz}}}, \bibinfo {author} {\bibfnamefont {V.}~\bibnamefont
			{{Vuleti{\'c}}}}, \ and\ \bibinfo {author} {\bibfnamefont {I.~L.}\
			\bibnamefont {{Chuang}}},\ }\href {\doibase 10.1103/PhysRevLett.103.103001}
	{\bibfield  {journal} {\bibinfo  {journal} {Physical Review Letters}\
		}\textbf {\bibinfo {volume} {103}},\ \bibinfo {eid} {103001} (\bibinfo {year}
		{2009})}\BibitemShut {NoStop}%
	\bibitem [{\citenamefont {{Romero-Isart}}\ \emph {et~al.}(2011)\citenamefont
		{{Romero-Isart}}, \citenamefont {{Pflanzer}}, \citenamefont {{Juan}},
		\citenamefont {{Quidant}}, \citenamefont {{Kiesel}}, \citenamefont
		{{Aspelmeyer}},\ and\ \citenamefont {{Cirac}}}]{Romero-Isart2011}%
	\BibitemOpen
	\bibfield  {author} {\bibinfo {author} {\bibfnamefont {O.}~\bibnamefont
			{{Romero-Isart}}}, \bibinfo {author} {\bibfnamefont {A.~C.}\ \bibnamefont
			{{Pflanzer}}}, \bibinfo {author} {\bibfnamefont {M.~L.}\ \bibnamefont
			{{Juan}}}, \bibinfo {author} {\bibfnamefont {R.}~\bibnamefont {{Quidant}}},
		\bibinfo {author} {\bibfnamefont {N.}~\bibnamefont {{Kiesel}}}, \bibinfo
		{author} {\bibfnamefont {M.}~\bibnamefont {{Aspelmeyer}}}, \ and\ \bibinfo
		{author} {\bibfnamefont {J.~I.}\ \bibnamefont {{Cirac}}},\ }\href {\doibase
		10.1103/PhysRevA.83.013803} {\bibfield  {journal} {\bibinfo  {journal}
			{\pra}\ }\textbf {\bibinfo {volume} {83}},\ \bibinfo {eid} {013803} (\bibinfo
		{year} {2011})}\BibitemShut {NoStop}%
	\bibitem [{\citenamefont {{Chang}}\ \emph {et~al.}(2010)\citenamefont
		{{Chang}}, \citenamefont {{Regal}}, \citenamefont {{Papp}}, \citenamefont
		{{Wilson}}, \citenamefont {{Ye}}, \citenamefont {{Painter}}, \citenamefont
		{{Kimble}},\ and\ \citenamefont {{Zoller}}}]{Chang2010}%
	\BibitemOpen
	\bibfield  {author} {\bibinfo {author} {\bibfnamefont {D.~E.}\ \bibnamefont
			{{Chang}}}, \bibinfo {author} {\bibfnamefont {C.~A.}\ \bibnamefont
			{{Regal}}}, \bibinfo {author} {\bibfnamefont {S.~B.}\ \bibnamefont {{Papp}}},
		\bibinfo {author} {\bibfnamefont {D.~J.}\ \bibnamefont {{Wilson}}}, \bibinfo
		{author} {\bibfnamefont {J.}~\bibnamefont {{Ye}}}, \bibinfo {author}
		{\bibfnamefont {O.}~\bibnamefont {{Painter}}}, \bibinfo {author}
		{\bibfnamefont {H.~J.}\ \bibnamefont {{Kimble}}}, \ and\ \bibinfo {author}
		{\bibfnamefont {P.}~\bibnamefont {{Zoller}}},\ }\href {\doibase
		10.1073/pnas.0912969107} {\bibfield  {journal} {\bibinfo  {journal}
			{Proceedings of the National Academy of Science}\ }\textbf {\bibinfo {volume}
			{107}},\ \bibinfo {pages} {1005} (\bibinfo {year} {2010})}\BibitemShut
	{NoStop}%
	\bibitem [{\citenamefont {{Nimmrichter}}\ \emph {et~al.}(2010)\citenamefont
		{{Nimmrichter}}, \citenamefont {{Hammerer}}, \citenamefont {{Asenbaum}},
		\citenamefont {{Ritsch}},\ and\ \citenamefont {{Arndt}}}]{Nimmrichter2010}%
	\BibitemOpen
	\bibfield  {author} {\bibinfo {author} {\bibfnamefont {S.}~\bibnamefont
			{{Nimmrichter}}}, \bibinfo {author} {\bibfnamefont {K.}~\bibnamefont
			{{Hammerer}}}, \bibinfo {author} {\bibfnamefont {P.}~\bibnamefont
			{{Asenbaum}}}, \bibinfo {author} {\bibfnamefont {H.}~\bibnamefont
			{{Ritsch}}}, \ and\ \bibinfo {author} {\bibfnamefont {M.}~\bibnamefont
			{{Arndt}}},\ }\href {\doibase 10.1088/1367-2630/12/8/083003} {\bibfield
		{journal} {\bibinfo  {journal} {New Journal of Physics}\ }\textbf {\bibinfo
			{volume} {12}},\ \bibinfo {eid} {083003} (\bibinfo {year}
		{2010})}\BibitemShut {NoStop}%
	\bibitem [{\citenamefont {{Russo}}\ \emph {et~al.}(2009)\citenamefont
		{{Russo}}, \citenamefont {{Barros}}, \citenamefont {{Stute}}, \citenamefont
		{{Dubin}}, \citenamefont {{Phillips}}, \citenamefont {{Monz}}, \citenamefont
		{{Northup}}, \citenamefont {{Becher}}, \citenamefont {{Salzburger}},
		\citenamefont {{Ritsch}}, \citenamefont {{Schmidt}},\ and\ \citenamefont
		{{Blatt}}}]{Russo2009}%
	\BibitemOpen
	\bibfield  {author} {\bibinfo {author} {\bibfnamefont {C.}~\bibnamefont
			{{Russo}}}, \bibinfo {author} {\bibfnamefont {H.~G.}\ \bibnamefont
			{{Barros}}}, \bibinfo {author} {\bibfnamefont {A.}~\bibnamefont {{Stute}}},
		\bibinfo {author} {\bibfnamefont {F.}~\bibnamefont {{Dubin}}}, \bibinfo
		{author} {\bibfnamefont {E.~S.}\ \bibnamefont {{Phillips}}}, \bibinfo
		{author} {\bibfnamefont {T.}~\bibnamefont {{Monz}}}, \bibinfo {author}
		{\bibfnamefont {T.~E.}\ \bibnamefont {{Northup}}}, \bibinfo {author}
		{\bibfnamefont {C.}~\bibnamefont {{Becher}}}, \bibinfo {author}
		{\bibfnamefont {T.}~\bibnamefont {{Salzburger}}}, \bibinfo {author}
		{\bibfnamefont {H.}~\bibnamefont {{Ritsch}}}, \bibinfo {author}
		{\bibfnamefont {P.~O.}\ \bibnamefont {{Schmidt}}}, \ and\ \bibinfo {author}
		{\bibfnamefont {R.}~\bibnamefont {{Blatt}}},\ }\href {\doibase
		10.1007/s00340-009-3430-2} {\bibfield  {journal} {\bibinfo  {journal}
			{Applied Physics B: Lasers and Optics}\ }\textbf {\bibinfo {volume} {95}},\
		\bibinfo {pages} {205} (\bibinfo {year} {2009})}\BibitemShut {NoStop}%
	\bibitem [{\citenamefont {{Cirac}}\ \emph {et~al.}(1992)\citenamefont
		{{Cirac}}, \citenamefont {{Blatt}}, \citenamefont {{Zoller}},\ and\
		\citenamefont {{Phillips}}}]{Cirac1992}%
	\BibitemOpen
	\bibfield  {author} {\bibinfo {author} {\bibfnamefont {J.~I.}\ \bibnamefont
			{{Cirac}}}, \bibinfo {author} {\bibfnamefont {R.}~\bibnamefont {{Blatt}}},
		\bibinfo {author} {\bibfnamefont {P.}~\bibnamefont {{Zoller}}}, \ and\
		\bibinfo {author} {\bibfnamefont {W.~D.}\ \bibnamefont {{Phillips}}},\ }\href
	{\doibase 10.1103/PhysRevA.46.2668} {\bibfield  {journal} {\bibinfo
			{journal} {\pra}\ }\textbf {\bibinfo {volume} {46}},\ \bibinfo {pages} {2668}
		(\bibinfo {year} {1992})}\BibitemShut {NoStop}%
	\bibitem [{\citenamefont {{Gieseler}}\ \emph {et~al.}(2012)\citenamefont
		{{Gieseler}}, \citenamefont {{Deutsch}}, \citenamefont {{Quidant}},\ and\
		\citenamefont {{Novotny}}}]{Gieseler2012}%
	\BibitemOpen
	\bibfield  {author} {\bibinfo {author} {\bibfnamefont {J.}~\bibnamefont
			{{Gieseler}}}, \bibinfo {author} {\bibfnamefont {B.}~\bibnamefont
			{{Deutsch}}}, \bibinfo {author} {\bibfnamefont {R.}~\bibnamefont
			{{Quidant}}}, \ and\ \bibinfo {author} {\bibfnamefont {L.}~\bibnamefont
			{{Novotny}}},\ }\href {\doibase 10.1103/PhysRevLett.109.103603} {\bibfield
		{journal} {\bibinfo  {journal} {Physical Review Letters}\ }\textbf {\bibinfo
			{volume} {109}},\ \bibinfo {eid} {103603} (\bibinfo {year}
		{2012})}\BibitemShut {NoStop}%
	\bibitem [{\citenamefont {{Nunnenkamp}}\ \emph {et~al.}(2010)\citenamefont
		{{Nunnenkamp}}, \citenamefont {{B{\o}rkje}}, \citenamefont {{Harris}},\ and\
		\citenamefont {{Girvin}}}]{Nunnenkamp2010}%
	\BibitemOpen
	\bibfield  {author} {\bibinfo {author} {\bibfnamefont {A.}~\bibnamefont
			{{Nunnenkamp}}}, \bibinfo {author} {\bibfnamefont {K.}~\bibnamefont
			{{B{\o}rkje}}}, \bibinfo {author} {\bibfnamefont {J.~G.~E.}\ \bibnamefont
			{{Harris}}}, \ and\ \bibinfo {author} {\bibfnamefont {S.~M.}\ \bibnamefont
			{{Girvin}}},\ }\href {\doibase 10.1103/PhysRevA.82.021806} {\bibfield
		{journal} {\bibinfo  {journal} {\pra}\ }\textbf {\bibinfo {volume} {82}},\
		\bibinfo {eid} {021806} (\bibinfo {year} {2010})}\BibitemShut {NoStop}%
	\bibitem [{\citenamefont {{Jain}}\ \emph {et~al.}(2016)\citenamefont {{Jain}},
		\citenamefont {{Gieseler}}, \citenamefont {{Moritz}}, \citenamefont
		{{Dellago}}, \citenamefont {{Quidant}},\ and\ \citenamefont
		{{Novotny}}}]{Jain2016}%
	\BibitemOpen
	\bibfield  {author} {\bibinfo {author} {\bibfnamefont {V.}~\bibnamefont
			{{Jain}}}, \bibinfo {author} {\bibfnamefont {J.}~\bibnamefont {{Gieseler}}},
		\bibinfo {author} {\bibfnamefont {C.}~\bibnamefont {{Moritz}}}, \bibinfo
		{author} {\bibfnamefont {C.}~\bibnamefont {{Dellago}}}, \bibinfo {author}
		{\bibfnamefont {R.}~\bibnamefont {{Quidant}}}, \ and\ \bibinfo {author}
		{\bibfnamefont {L.}~\bibnamefont {{Novotny}}},\ }\href {\doibase
		10.1103/PhysRevLett.116.243601} {\bibfield  {journal} {\bibinfo  {journal}
			{Physical Review Letters}\ }\textbf {\bibinfo {volume} {116}},\ \bibinfo
		{eid} {243601} (\bibinfo {year} {2016})}\BibitemShut {NoStop}%
	\bibitem [{\citenamefont {{Genes}}\ \emph {et~al.}(2008)\citenamefont
		{{Genes}}, \citenamefont {{Mari}}, \citenamefont {{Tombesi}},\ and\
		\citenamefont {{Vitali}}}]{Genes2008Entanglement}%
	\BibitemOpen
	\bibfield  {author} {\bibinfo {author} {\bibfnamefont {C.}~\bibnamefont
			{{Genes}}}, \bibinfo {author} {\bibfnamefont {A.}~\bibnamefont {{Mari}}},
		\bibinfo {author} {\bibfnamefont {P.}~\bibnamefont {{Tombesi}}}, \ and\
		\bibinfo {author} {\bibfnamefont {D.}~\bibnamefont {{Vitali}}},\ }\href
	{\doibase 10.1103/PhysRevA.78.032316} {\bibfield  {journal} {\bibinfo
			{journal} {\pra}\ }\textbf {\bibinfo {volume} {78}},\ \bibinfo {eid} {032316}
		(\bibinfo {year} {2008})}\BibitemShut {NoStop}%
	\bibitem [{\citenamefont {{Hofer}}(2015)}]{HoferPhD}%
	\BibitemOpen
	\bibfield  {author} {\bibinfo {author} {\bibfnamefont {S.~G.}\ \bibnamefont
			{{Hofer}}},\ }\href@noop {} {\emph {\bibinfo {title} {{Quantum Control of
					Optomechanical Systems}, PhD Thesis}}}\ (\bibinfo  {publisher} {University of
		Vienna},\ \bibinfo {year} {2015})\BibitemShut {NoStop}%
	\bibitem [{\citenamefont {{Windey}}\ \emph {et~al.}(2018)\citenamefont
		{{Windey}}, \citenamefont {{Gonzalez-Ballestero}}, \citenamefont {{Maurer}},
		\citenamefont {{Novotny}}, \citenamefont {{Romero-Isart}},\ and\
		\citenamefont {{Reimann}}}]{Windey2018}%
	\BibitemOpen
	\bibfield  {author} {\bibinfo {author} {\bibfnamefont {D.}~\bibnamefont
			{{Windey}}}, \bibinfo {author} {\bibfnamefont {C.}~\bibnamefont
			{{Gonzalez-Ballestero}}}, \bibinfo {author} {\bibfnamefont {P.}~\bibnamefont
			{{Maurer}}}, \bibinfo {author} {\bibfnamefont {L.}~\bibnamefont {{Novotny}}},
		\bibinfo {author} {\bibfnamefont {O.}~\bibnamefont {{Romero-Isart}}}, \ and\
		\bibinfo {author} {\bibfnamefont {R.}~\bibnamefont {{Reimann}}},\ }\href@noop
	{} {\bibfield  {journal} {\bibinfo  {journal} {arXiv e-prints}\ } (\bibinfo
		{year} {2018})},\ \Eprint {http://arxiv.org/abs/1812.09176} {arXiv:1812.09176
		[quant-ph]} \BibitemShut {NoStop}%
	\bibitem [{\citenamefont {{Gonzalez-Ballestero}}\ \emph
		{et~al.}(2019)\citenamefont {{Gonzalez-Ballestero}}, \citenamefont
		{{Maurer}}, \citenamefont {{Windey}}, \citenamefont {{Novotny}},
		\citenamefont {{Reimann}},\ and\ \citenamefont
		{{Romero-Isart}}}]{Gonzalez2019}%
	\BibitemOpen
	\bibfield  {author} {\bibinfo {author} {\bibfnamefont {C.}~\bibnamefont
			{{Gonzalez-Ballestero}}}, \bibinfo {author} {\bibfnamefont {P.}~\bibnamefont
			{{Maurer}}}, \bibinfo {author} {\bibfnamefont {D.}~\bibnamefont {{Windey}}},
		\bibinfo {author} {\bibfnamefont {L.}~\bibnamefont {{Novotny}}}, \bibinfo
		{author} {\bibfnamefont {R.}~\bibnamefont {{Reimann}}}, \ and\ \bibinfo
		{author} {\bibfnamefont {O.}~\bibnamefont {{Romero-Isart}}},\ }\href@noop {}
	{\bibfield  {journal} {\bibinfo  {journal} {arXiv e-prints}\ } (\bibinfo
		{year} {2019})},\ \Eprint {http://arxiv.org/abs/1902.01282} {arXiv:1902.01282
		[quant-ph]} \BibitemShut {NoStop}%
	
\setcounter{firstbib}{\value{enumiv}}
\end{thebibliography}

\begin{thebibliography}{10}%
	\makeatletter
	\providecommand \@ifxundefined [1]{%
		\@ifx{#1\undefined}
	}%
	\providecommand \@ifnum [1]{%
		\ifnum #1\expandafter \@firstoftwo
		\else \expandafter \@secondoftwo
		\fi
	}%
	\providecommand \@ifx [1]{%
		\ifx #1\expandafter \@firstoftwo
		\else \expandafter \@secondoftwo
		\fi
	}%
	\providecommand \natexlab [1]{#1}%
	\providecommand \enquote  [1]{``#1''}%
	\providecommand \bibnamefont  [1]{#1}%
	\providecommand \bibfnamefont [1]{#1}%
	\providecommand \citenamefont [1]{#1}%
	\providecommand \href@noop [0]{\@secondoftwo}%
	\providecommand \href [0]{\begingroup \@sanitize@url \@href}%
	\providecommand \@href[1]{\@@startlink{#1}\@@href}%
	\providecommand \@@href[1]{\endgroup#1\@@endlink}%
	\providecommand \@sanitize@url [0]{\catcode `\\12\catcode `\$12\catcode
		`\&12\catcode `\#12\catcode `\^12\catcode `\_12\catcode `\%12\relax}%
	\providecommand \@@startlink[1]{}%
	\providecommand \@@endlink[0]{}%
	\providecommand \url  [0]{\begingroup\@sanitize@url \@url }%
	\providecommand \@url [1]{\endgroup\@href {#1}{\urlprefix }}%
	\providecommand \urlprefix  [0]{URL }%
	\providecommand \Eprint [0]{\href }%
	\providecommand \doibase [0]{http://dx.doi.org/}%
	\providecommand \selectlanguage [0]{\@gobble}%
	\providecommand \bibinfo  [0]{\@secondoftwo}%
	\providecommand \bibfield  [0]{\@secondoftwo}%
	\providecommand \translation [1]{[#1]}%
	\providecommand \BibitemOpen [0]{}%
	\providecommand \bibitemStop [0]{}%
	\providecommand \bibitemNoStop [0]{.\EOS\space}%
	\providecommand \EOS [0]{\spacefactor3000\relax}%
	\providecommand \BibitemShut  [1]{\csname bibitem#1\endcsname}%
	\let\auto@bib@innerbib\@empty
	%</preamble>
	\bibitem [{\citenamefont {{Novotny}}\ and\ \citenamefont
		{{Hecht}}(2012)}]{NovotnyBook}%
	\BibitemOpen
	\bibfield  {author} {\bibinfo {author} {\bibfnamefont {L.}~\bibnamefont
			{{Novotny}}}\ and\ \bibinfo {author} {\bibfnamefont {B.}~\bibnamefont
			{{Hecht}}},\ }\href@noop {} {\emph {\bibinfo {title} {Principles of
				Nano-Optics, by Lukas Novotny , Bert Hecht, Cambridge, UK: Cambridge
				University Press, 2012}}}\ (\bibinfo {year} {2012})\BibitemShut {NoStop}%
	\bibitem [{\citenamefont {{Tanji-Suzuki}}\ \emph {et~al.}(2011)\citenamefont
		{{Tanji-Suzuki}}, \citenamefont {{Leroux}}, \citenamefont {{Schleier-Smith}},
		\citenamefont {{Cetina}}, \citenamefont {{Grier}}, \citenamefont {{Simon}},\
		and\ \citenamefont {{Vuleti{\'c}}}}]{Tanji-Suzuki2011}%
	\BibitemOpen
	\bibfield  {author} {\bibinfo {author} {\bibfnamefont {H.}~\bibnamefont
			{{Tanji-Suzuki}}}, \bibinfo {author} {\bibfnamefont {I.~D.}\ \bibnamefont
			{{Leroux}}}, \bibinfo {author} {\bibfnamefont {M.~H.}\ \bibnamefont
			{{Schleier-Smith}}}, \bibinfo {author} {\bibfnamefont {M.}~\bibnamefont
			{{Cetina}}}, \bibinfo {author} {\bibfnamefont {A.~T.}\ \bibnamefont
			{{Grier}}}, \bibinfo {author} {\bibfnamefont {J.}~\bibnamefont {{Simon}}}, \
		and\ \bibinfo {author} {\bibfnamefont {V.}~\bibnamefont {{Vuleti{\'c}}}},\
	}\href {\doibase 10.1016/B978-0-12-385508-4.00004-8} {\bibfield  {journal}
		{\bibinfo  {journal} {Advances in Atomic Molecular and Optical Physics}\
		}\textbf {\bibinfo {volume} {60}},\ \bibinfo {pages} {201} (\bibinfo {year}
		{2011})},\ \Eprint {http://arxiv.org/abs/1104.3594} {arXiv:1104.3594
		[quant-ph]} \BibitemShut {NoStop}%
	\bibitem [{\citenamefont {{Motsch}}\ \emph {et~al.}(2010)\citenamefont
		{{Motsch}}, \citenamefont {{Zeppenfeld}}, \citenamefont {{Pinkse}},\ and\
		\citenamefont {{Rempe}}}]{Motsch2010}%
	\BibitemOpen
	\bibfield  {author} {\bibinfo {author} {\bibfnamefont {M.}~\bibnamefont
			{{Motsch}}}, \bibinfo {author} {\bibfnamefont {M.}~\bibnamefont
			{{Zeppenfeld}}}, \bibinfo {author} {\bibfnamefont {P.~W.~H.}\ \bibnamefont
			{{Pinkse}}}, \ and\ \bibinfo {author} {\bibfnamefont {G.}~\bibnamefont
			{{Rempe}}},\ }\href {\doibase 10.1088/1367-2630/12/6/063022} {\bibfield
		{journal} {\bibinfo  {journal} {New Journal of Physics}\ }\textbf {\bibinfo
			{volume} {12}},\ \bibinfo {eid} {063022} (\bibinfo {year} {2010})},\ \Eprint
	{http://arxiv.org/abs/0904.4405} {arXiv:0904.4405 [quant-ph]} \BibitemShut
	{NoStop}%
	\bibitem [{\citenamefont {{Delic}}\ \emph {et~al.}()\citenamefont
		{{Deli{\'c}}}, \citenamefont {{Grass}}, \citenamefont {{Reisenbauer}},
		\citenamefont {{Kiesel}},\ and\ \citenamefont {{Aspelmeyer}}}]{Delicb2018}%
	\BibitemOpen
	\bibfield  {author} {\bibinfo {author} {\bibfnamefont {U.}~\bibnamefont
			{{Deli{\'c}}}}, \bibinfo {author} {\bibfnamefont {D.}~\bibnamefont
			{{Grass}}}, \bibinfo {author} {\bibfnamefont {M.}~\bibnamefont
			{{Reisenbauer}}}, \bibinfo {author} {\bibfnamefont {N.}~\bibnamefont
			{{Kiesel}}}, \ and\ \bibinfo {author} {\bibfnamefont {M.}~\bibnamefont
			{{Aspelmeyer}}},\ }\href@noop {} {\bibinfo  {journal} {(unpublished)}\
	}\BibitemShut {NoStop}%
	\bibitem [{\citenamefont {{Genes}}\ \emph {et~al.}(2008)\citenamefont
		{{Genes}}, \citenamefont {{Vitali}}, \citenamefont {{Tombesi}}, \citenamefont
		{{Gigan}},\ and\ \citenamefont {{Aspelmeyer}}}]{Genes2008}%
	\BibitemOpen
	\bibfield  {journal} {  }\bibfield  {author} {\bibinfo {author} {\bibfnamefont
			{C.}~\bibnamefont {{Genes}}}, \bibinfo {author} {\bibfnamefont
			{D.}~\bibnamefont {{Vitali}}}, \bibinfo {author} {\bibfnamefont
			{P.}~\bibnamefont {{Tombesi}}}, \bibinfo {author} {\bibfnamefont
			{S.}~\bibnamefont {{Gigan}}}, \ and\ \bibinfo {author} {\bibfnamefont
			{M.}~\bibnamefont {{Aspelmeyer}}},\ }\href {\doibase
		10.1103/PhysRevA.77.033804} {\bibfield  {journal} {\bibinfo  {journal}
			{\pra}\ }\textbf {\bibinfo {volume} {77}},\ \bibinfo {eid} {033804} (\bibinfo
		{year} {2008})},\ \Eprint {http://arxiv.org/abs/0705.1728} {arXiv:0705.1728
		[quant-ph]} \BibitemShut {NoStop}%
	\bibitem [{\citenamefont {{Safavi-Naeini}}\ \emph {et~al.}(2013)\citenamefont
		{{Safavi-Naeini}}, \citenamefont {{Chan}}, \citenamefont {{Hill}},
		\citenamefont {{Gr{\"o}blacher}}, \citenamefont {{Miao}}, \citenamefont
		{{Chen}}, \citenamefont {{Aspelmeyer}},\ and\ \citenamefont
		{{Painter}}}]{Safavi-Naeinib2013}%
	\BibitemOpen
	\bibfield  {author} {\bibinfo {author} {\bibfnamefont {A.~H.}\ \bibnamefont
			{{Safavi-Naeini}}}, \bibinfo {author} {\bibfnamefont {J.}~\bibnamefont
			{{Chan}}}, \bibinfo {author} {\bibfnamefont {J.~T.}\ \bibnamefont {{Hill}}},
		\bibinfo {author} {\bibfnamefont {S.}~\bibnamefont {{Gr{\"o}blacher}}},
		\bibinfo {author} {\bibfnamefont {H.}~\bibnamefont {{Miao}}}, \bibinfo
		{author} {\bibfnamefont {Y.}~\bibnamefont {{Chen}}}, \bibinfo {author}
		{\bibfnamefont {M.}~\bibnamefont {{Aspelmeyer}}}, \ and\ \bibinfo {author}
		{\bibfnamefont {O.}~\bibnamefont {{Painter}}},\ }\href {\doibase
		10.1088/1367-2630/15/3/035007} {\bibfield  {journal} {\bibinfo  {journal}
			{New Journal of Physics}\ }\textbf {\bibinfo {volume} {15}},\ \bibinfo {eid}
		{035007} (\bibinfo {year} {2013})},\ \Eprint {http://arxiv.org/abs/1210.2671}
	{arXiv:1210.2671 [physics.optics]} \BibitemShut {NoStop}%
	\bibitem [{\citenamefont {{Rabl}}\ \emph {et~al.}(2009)\citenamefont {{Rabl}},
		\citenamefont {{Genes}}, \citenamefont {{Hammerer}},\ and\ \citenamefont
		{{Aspelmeyer}}}]{Rablb2009}%
	\BibitemOpen
	\bibfield  {author} {\bibinfo {author} {\bibfnamefont {P.}~\bibnamefont
			{{Rabl}}}, \bibinfo {author} {\bibfnamefont {C.}~\bibnamefont {{Genes}}},
		\bibinfo {author} {\bibfnamefont {K.}~\bibnamefont {{Hammerer}}}, \ and\
		\bibinfo {author} {\bibfnamefont {M.}~\bibnamefont {{Aspelmeyer}}},\ }\href
	{\doibase 10.1103/PhysRevA.80.063819} {\bibfield  {journal} {\bibinfo
			{journal} {\pra}\ }\textbf {\bibinfo {volume} {80}},\ \bibinfo {eid} {063819}
		(\bibinfo {year} {2009})},\ \Eprint {http://arxiv.org/abs/0903.1637}
	{arXiv:0903.1637 [quant-ph]} \BibitemShut {NoStop}%
	\bibitem [{\citenamefont {{Jayich}}\ \emph {et~al.}(2012)\citenamefont
		{{Jayich}}, \citenamefont {{Sankey}}, \citenamefont {{B{\o}rkje}},
		\citenamefont {{Lee}}, \citenamefont {{Yang}}, \citenamefont {{Underwood}},
		\citenamefont {{Childress}}, \citenamefont {{Petrenko}}, \citenamefont
		{{Girvin}},\ and\ \citenamefont {{Harris}}}]{Jayichb2012}%
	\BibitemOpen
	\bibfield  {author} {\bibinfo {author} {\bibfnamefont {A.~M.}\ \bibnamefont
			{{Jayich}}}, \bibinfo {author} {\bibfnamefont {J.~C.}\ \bibnamefont
			{{Sankey}}}, \bibinfo {author} {\bibfnamefont {K.}~\bibnamefont
			{{B{\o}rkje}}}, \bibinfo {author} {\bibfnamefont {D.}~\bibnamefont {{Lee}}},
		\bibinfo {author} {\bibfnamefont {C.}~\bibnamefont {{Yang}}}, \bibinfo
		{author} {\bibfnamefont {M.}~\bibnamefont {{Underwood}}}, \bibinfo {author}
		{\bibfnamefont {L.}~\bibnamefont {{Childress}}}, \bibinfo {author}
		{\bibfnamefont {A.}~\bibnamefont {{Petrenko}}}, \bibinfo {author}
		{\bibfnamefont {S.~M.}\ \bibnamefont {{Girvin}}}, \ and\ \bibinfo {author}
		{\bibfnamefont {J.~G.~E.}\ \bibnamefont {{Harris}}},\ }\href {\doibase
		10.1088/1367-2630/14/11/115018} {\bibfield  {journal} {\bibinfo  {journal}
			{New Journal of Physics}\ }\textbf {\bibinfo {volume} {14}},\ \bibinfo {eid}
		{115018} (\bibinfo {year} {2012})},\ \Eprint {http://arxiv.org/abs/1209.2730}
	{arXiv:1209.2730 [physics.optics]} \BibitemShut {NoStop}%
	\bibitem [{\citenamefont {{Jain}}\ \emph {et~al.}(2016)\citenamefont {{Jain}},
		\citenamefont {{Gieseler}}, \citenamefont {{Moritz}}, \citenamefont
		{{Dellago}}, \citenamefont {{Quidant}},\ and\ \citenamefont
		{{Novotny}}}]{Jainb2016}%
	\BibitemOpen
	\bibfield  {author} {\bibinfo {author} {\bibfnamefont {V.}~\bibnamefont
			{{Jain}}}, \bibinfo {author} {\bibfnamefont {J.}~\bibnamefont {{Gieseler}}},
		\bibinfo {author} {\bibfnamefont {C.}~\bibnamefont {{Moritz}}}, \bibinfo
		{author} {\bibfnamefont {C.}~\bibnamefont {{Dellago}}}, \bibinfo {author}
		{\bibfnamefont {R.}~\bibnamefont {{Quidant}}}, \ and\ \bibinfo {author}
		{\bibfnamefont {L.}~\bibnamefont {{Novotny}}},\ }\href {\doibase
		10.1103/PhysRevLett.116.243601} {\bibfield  {journal} {\bibinfo  {journal}
			{Physical Review Letters}\ }\textbf {\bibinfo {volume} {116}},\ \bibinfo
		{eid} {243601} (\bibinfo {year} {2016})},\ \Eprint
	{http://arxiv.org/abs/1603.03420} {arXiv:1603.03420 [physics.optics]}
	\BibitemShut {NoStop}%
	\bibitem [{\citenamefont {{Nunnenkamp}}\ \emph {et~al.}(2010)\citenamefont
		{{Nunnenkamp}}, \citenamefont {{B{\o}rkje}}, \citenamefont {{Harris}},\ and\
		\citenamefont {{Girvin}}}]{Nunnenkampb2010}%
	\BibitemOpen
	\bibfield  {author} {\bibinfo {author} {\bibfnamefont {A.}~\bibnamefont
			{{Nunnenkamp}}}, \bibinfo {author} {\bibfnamefont {K.}~\bibnamefont
			{{B{\o}rkje}}}, \bibinfo {author} {\bibfnamefont {J.~G.~E.}\ \bibnamefont
			{{Harris}}}, \ and\ \bibinfo {author} {\bibfnamefont {S.~M.}\ \bibnamefont
			{{Girvin}}},\ }\href {\doibase 10.1103/PhysRevA.82.021806} {\bibfield
		{journal} {\bibinfo  {journal} {\pra}\ }\textbf {\bibinfo {volume} {82}},\
		\bibinfo {eid} {021806} (\bibinfo {year} {2010})},\ \Eprint
	{http://arxiv.org/abs/1004.2510} {arXiv:1004.2510 [cond-mat.mes-hall]}
	\BibitemShut {NoStop}%
\end{thebibliography}
\end{document}